\documentclass[review]{elsarticle}
\usepackage[utf8]{inputenc}

\usepackage{graphicx} \graphicspath{{pics/}}
\usepackage{amsmath,amssymb}
\usepackage{textcomp} 
\usepackage{gensymb}
\usepackage{url}
\usepackage{siunitx}

\usepackage{tabularx}

\usepackage{floatrow}
\usepackage[font=small,labelfont=bf]{caption}

\usepackage{soul}

\usepackage{algorithm}
\usepackage{algpseudocode}

\usepackage[nottoc,numbib]{tocbibind} 

\usepackage{geometry} \newgeometry{hmargin={20mm,20mm}}   

\journal{arXiv}

\begin{document}

\begin{frontmatter}

\title{Directed energy deposition powder stream modeling using a Gaussian beam ray representation}

\author[1]{A. Martinez-Marchese\corref{mycorrespondingauthor}}
\cortext[mycorrespondingauthor]{Corresponding author}
\ead{alex.martinez@uwaterloo.ca}

\author[2]{M. Klumpp}
\author[1]{E. Toyserkani}

\address[1]{Multi-Scale Additive Manufacturing (MSAM) Lab, Department of Mechanical and Mechatronics Engineering, University of Waterloo, Waterloo, N2L 3G1, ON, Canada}

\address[2]{Institute of Analysis, Dynamics and Modeling, University of Stuttgart, Pfaffenwaldring 57, 70569 Stuttgart, Germany}

\begin{abstract}
The powder stream from a side feed nozzle, or part of the powder stream in some coaxial nozzles, in a directed energy deposition via powder feeding (DED-PF) machine, can be modeled using a particle velocity field that has a constant downward component and a linearly increasing outward component, in proportion to the powder stream's center line distance \cite{martinezmarchese2022application}. However, when the powder stream is subject to a force field, it was found that the shape of the powder concentration function close to the center of the powder stream diverges considerably at high degrees of focusing. This discrepancy is reduced by modeling the powder stream based on the ray representation of a Gaussian beam \cite{pang2021modeling}. Experimental results from high-speed camera particle tracking and numerically extrapolating the trajectories to the nozzle exit suggests that the statistics of the powder stream correspond to this model. A novel method to compute the particle concentration along the stream using an optical system analog, with the focusing force field modeled as the transfer matrix of a graded refractive index (GRIN) lens, is also demonstrated. This method is orders of magnitude faster than the corresponding Lagrangian simulation.
\end{abstract}


\end{frontmatter}

\section{Introduction}
The powder motion models found in the directed energy deposition via powder feeding (DED-PF) literature are either numerical simulations taking into account nozzle geometry, gas flow and particle motion, or analytic models where the powder stream's powder concentration distribution (PCD) is static.

Analytic models usually formulate the PCD using a Gaussian function that has a linearly increasing intensity ratio width (IRW) that matches experimental data \cite{huang2016comprehensive, haley2019working}. These types of models are orders of magnitude faster to compute than numerical models, however the interaction of the powder stream with forces due to shielding gas \cite{toyserkani2004laser} or reflections with the build plate are not taken into account or are implicitly taken into account by empirically fitting the model to a measured PCD in a particular DED-PF setup. In this article, a third type of model is presented; one that has a computationally efficient analytic solution, but is also able to account for forces being applied to the powder stream. Note that the scope of models in the literature being considered and the one in this article are for the powder motion downstream of the nozzle exit.

The mechanism causing linearly increasing IRW in analytic models is usually explained by some diffusion mechanism that reproduces the observed linearly increasing IRW \cite{lin1999concentration, fuks1989mechanics}. This mechanism is plausible if one assumes there are a high number of particle collisions in the nozzle exit zone producing some effective diffusion of the particle concentration. In \cite{martinezmarchese2022application}, by observing high-speed camera footage of a powder stream, it was determined that the number of collisions is low for DED-PF relevant process parameters. A more plausible mechanism was then proposed, based on a constant downward component and a linearly increasing outward component, with the outward component being proportional to the distance from the powder stream's center line. This powder motion model was also used to accurately predict the behaviour of the PCD downstream of an externally applied sound radiation focusing force field to a powder stream.

\begin{figure}[!ht]
\center
\vspace{0cm} 
\includegraphics[width=15cm]{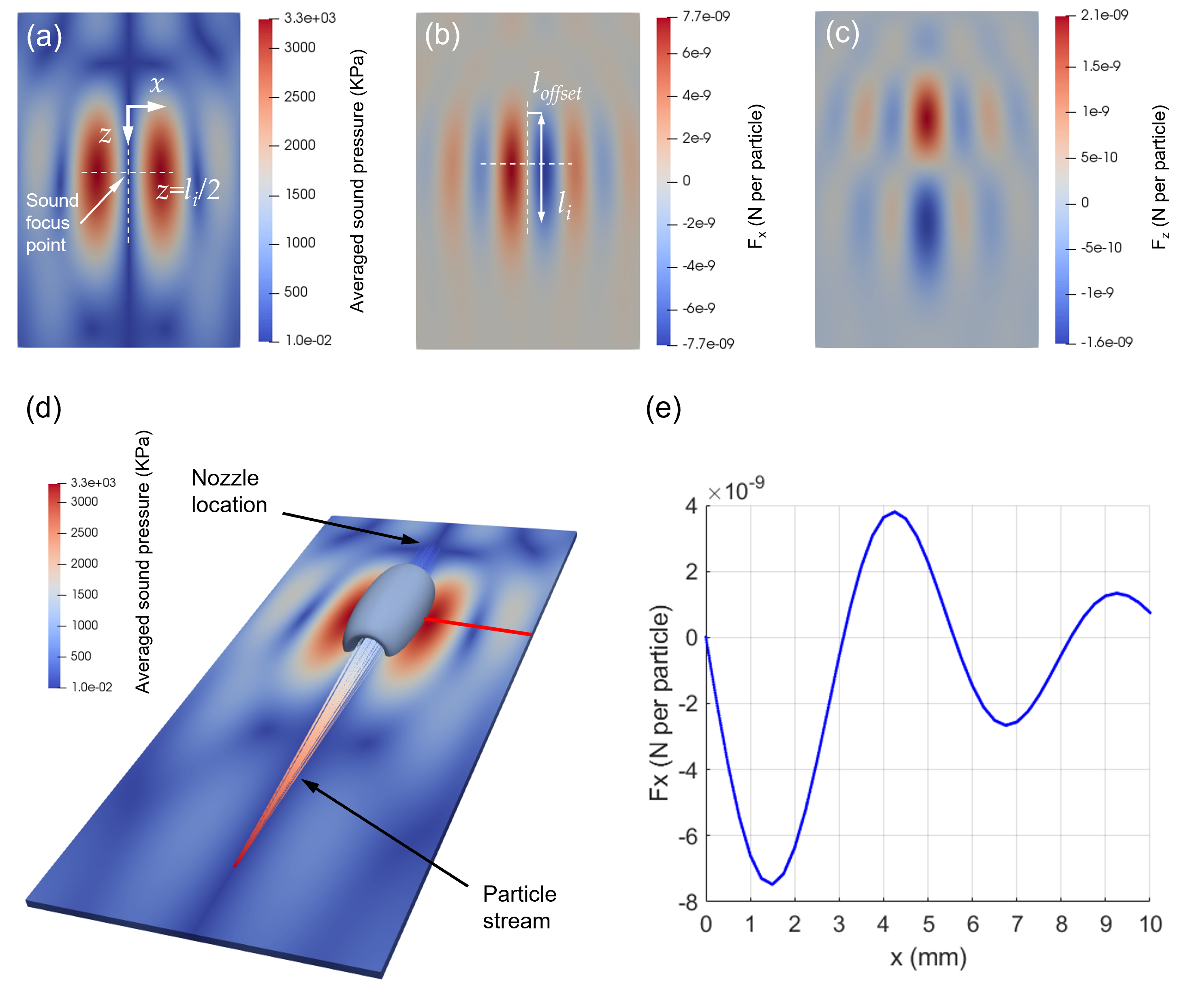}\\ 
\caption{Period averaged sound pressure (ASP) (a), x component of the resulting force field (b), z component of the resulting force field (c). ASP field, with particles tracks, and isosurface with a radial sound radiation force value of \newline-4.0\ignorespaces$\times 10^{-9}$ N per particle, based on the setup for SS 316L particles with sound produced at 16V \cite{martinezmarchese2022application} (d), sound radiation force in the x-axis direction along the red line (e)}
\vspace{0.5cm}
\label{fig:vortexFieldProps}
\end{figure}

In particular, the focusing force field considered in \cite{martinezmarchese2022application} is a result of high power sound waves coming for the area surrounding the powder stream, interacting with the particles and producing a force field that permeates the powder stream and accelerates particles depending on their position in the field. Plots of the sound pressure and the resulting components of this particular force field for the case of SS 316L particles with sound produced by applying 16V to ultrasound transducers in the setup described in \cite{martinezmarchese2022application} are shown in Fig.~\ref{fig:vortexFieldProps}. The variables shown in the figure are explained in Section~\ref{sec:abcd}. The transducers form an array with the shape of the hemisphere with an open top for the powder to move through it, with the center of the sphere aligned with the sound focus point. Note that the methods described in this article may also be used to quickly model other types of sound fields being applied, such as force fields due to electromagnetic fields \cite{huang2019electrodynamic}, as long as the powder stream can be modeled using a Gaussian beam as described in the rest of this article.

\begin{figure}[!ht]
\center
\vspace{0cm} 
\includegraphics[width=12cm]{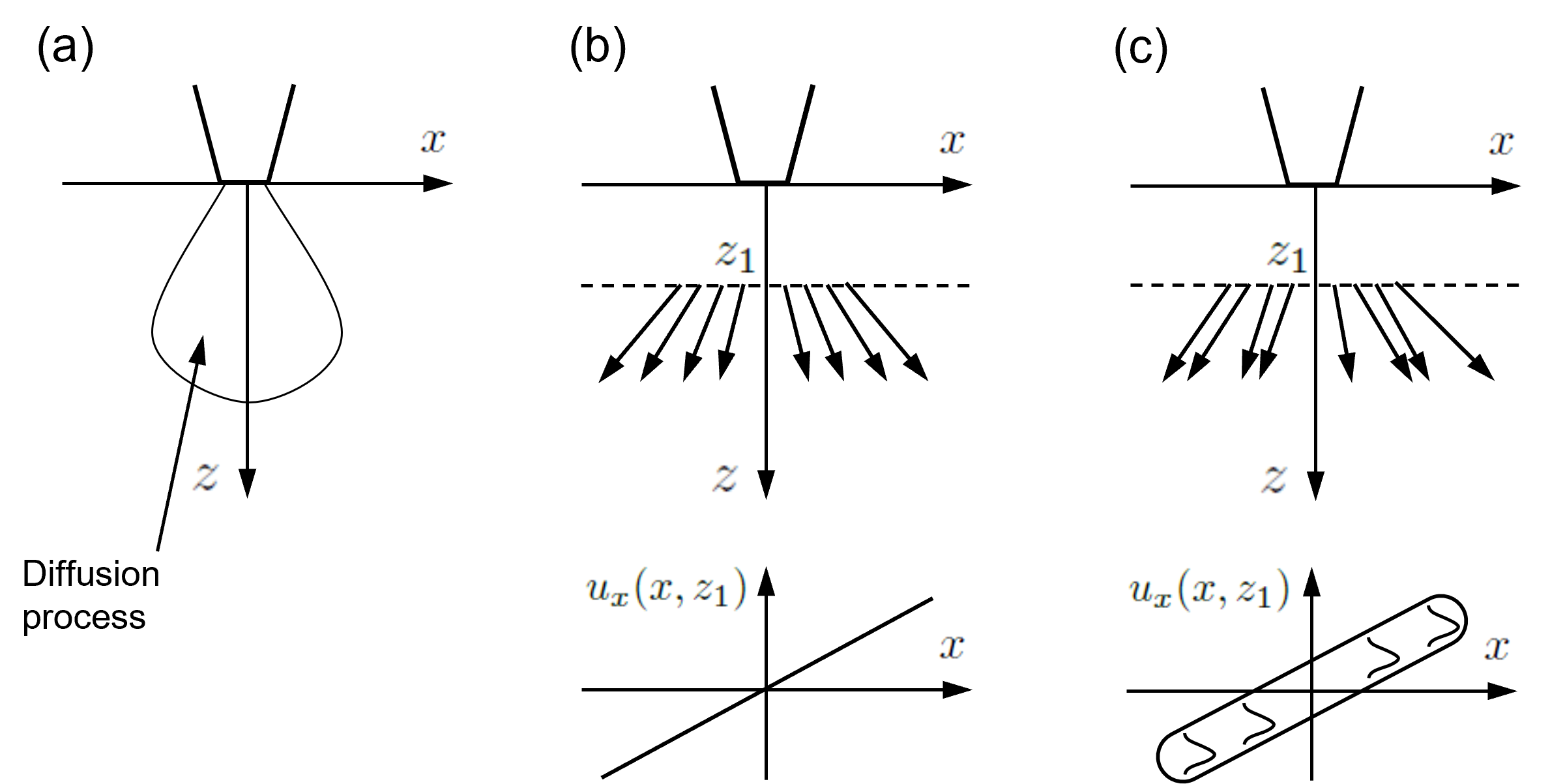}\\ 
\caption{Particle motion mechanisms; (a) diffusion \cite{lin1999concentration}, (b) initial velocities with increasing tangential components \cite{martinezmarchese2022application}, (c) hypothesis in this article}
\vspace{0.5cm}
\label{fig:particleMotion}
\end{figure}

For high degrees of focusing, the powder motion model in \cite{martinezmarchese2022application} shows a higher peak concentration compared to the experimental results, that approach a constant concentration and IRW. A possible explanation for this discrepancy is that besides a linearly increasing outward particle speed, there is also a distribution of possible deviations from this speed. This would introduce an `aberration' in the focused particles that would produce a less focused powder stream for the same applied force field. A summary of the mechanisms used to explain the behaviour of the powder steam for the DED-PF process is shown in Fig.~\ref{fig:particleMotion}.

\begin{figure}[!ht]
\center
\vspace{0.2cm} 
\includegraphics[width=10cm]{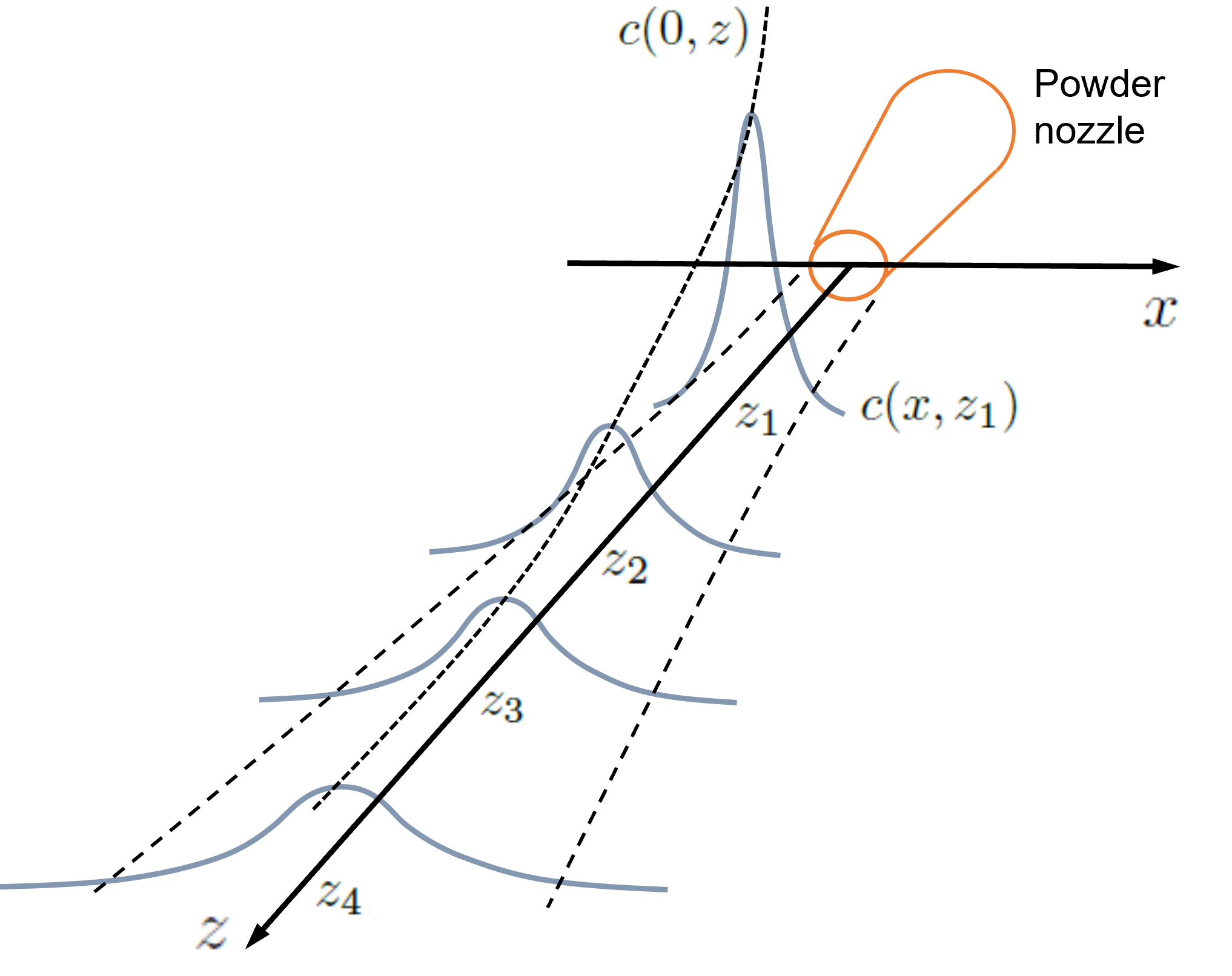}\\ 
\caption{Powder PCD and some of its properties: symmetry about z, mass conservation (polar integral about $z$of $c(x,z)$ equal for all $z$-values), IRW, given by value of powder radius $r$ such that $c(r,z_n)/c(0,z_n) = k$ (where $k$ is a constant such as $e^{-1}$ or $e^{-2}$) increases linearly with $z$ downstream of the nozzle}
\vspace{0.2cm}
\label{fig:powderStreamProperties}
\end{figure}

The function used to describe the powder flow rate for a DED-PF nozzle in the literature is given by \cite{huang2016comprehensive}:

\begin{equation} \label{eq:samsol}
c = \frac{\dot{m}}{U m_p \pi r(z)^2} \exp\left(- \frac{r^2}{r(z)^2}\right)
\end{equation}
\hfill \break
where $\dot{m}$ is the mass flowrate, $U$ is the particle speed approximately equal to the downward speed $u_{0z}$, and $m_p$ is the particle mass. The effective powder radius given by:

\begin{equation} \label{eq:RzSam}
r(z) = r_0 + \tan (\theta) z
\end{equation}
\hfill \break
\noindent Some properties of the PCD based on Eqs.~\ref{eq:samsol} and \ref{eq:RzSam} are shown in Fig.~\ref{fig:powderStreamProperties}

Integrating Eq.~\ref{eq:samsol} in polar coordinates results in the number of particles crossing a plane perpendicular to the powder stream axis. As expected, this value is constant due to mass conservation, and proportional to the mass flow rate. The value is given by:

\begin{equation} \label{eq:np}
n_p = \frac{\dot{m}}{U m_p}
\end{equation}
\hfill \break
The mechanism that produces the PCD described by Eq.~\ref{eq:samsol} can be derived from a diffusion equation \cite{lin1999concentration}. This is a fair approximation for small particles \cite{fuks1989mechanics}, where some particle collisions close to the nozzle might be taking place that can be modeled by a diffusion process. However, as seen in the high-speed video data in \cite{martinezmarchese2022application}, most particles are not colliding, especially away from the nozzle, and are seen to be spreading outwards. In the same study, to accurately couple the powder stream to a force field, a new mechanism was derived, in which an initial velocity field that reproduces Eq.~\ref{eq:samsol} without a diffusion mechanism for the particles is given by:

\begin{equation} \label{eq:alexSol}
u_r(r,0) = U \tan (\theta) \frac{r}{r_0} 
\end{equation}
\hfill \break
This was done by finding an initial particle speed distribution for the Eulerian continuous concentration distribution equation given in \cite{fuks1989mechanics} and referenced by Lin in his original derivation of Eq.~\ref{eq:samsol} \cite{lin1999concentration}, but without a diffusion term, such that the same solution is recovered. In this article, it will be shown that a model producing a linear dependence between tangential speed and $x$-position with some offset due to a probability distribution as shown in Fig.~\ref{fig:particleMotion} (c) corresponds to considering the particle tracks as rays and having the same spatial and angular probability distribution that is used in some models of laser beam propagation \cite{ pang2021modeling, crooker2006representation, milsom2000ray, colbourne2019representation, bastiaans1978wigner}.

A laser beam can be described via the Helmholtz wave equation or the Fresnel-Kirchhoff integral \cite{svelto1998principles, siegman1986lasers}. One can solve for the electric field of a spherical wave in the paraxial approximation for a beam that has a small angle with respect to one of the coordinate axes. However, the solution using a real number on an axis for the source location produces a nonphysical electric field because its amplitude does not fall off quickly enough in the transverse direction, and therefore carries infinite energy in a transverse plane \cite{siegman1986lasers}. One can obtain a more physical solution for the electric field by using a complex source position, producing a real Gaussian factor hence the name Gaussian beam. This also corresponds to the electric field of a freely propagating laser beam. For a beam along the $z$-axis, this field (normalized to an amplitude of one) is given by \cite{svelto1998principles}:

\begin{equation} \label{eq:laserEfield}
E(x,y,z) = \frac{w_0}{w(z)} \exp\left(-\frac{x^2 + y^2}{w(z)^2}+j\phi\right)
\end{equation} 
\hfill \break
with $j = \sqrt{-1}$, $\phi = \arctan (z/z_R) - k (x^2 + y^2) / (2R)$ and $k = 2 \pi / \lambda$.
The functions for the beam half width $w$ and the beam radius of curvature $R$ can be found from Eqs.~\ref{eq:w} and \ref{eq:R}.

\begin{align} 
w(z) &= w_0 \sqrt{1 + \frac{z^2}{z_R^2}} \label{eq:w} \\
R(z) &= z + \frac{z_R^2}{z} \label{eq:R}
\end{align} 
\hfill \break
where $w_0$ is the beam half width. The Rayleigh range, $z_R$, is given by:

\begin{equation} \label{eq:zR}
z_R = \frac{\pi w_0^2}{\lambda}
\end{equation} 
\hfill \break
where $\lambda$ is the wavelength. This free space propagation solution uses the complex source position $q$ given by:

\begin{equation} \label{eq:complexQ}
\frac{1}{q} = \frac{1}{R} - j \frac{\lambda}{\pi w^2}
\end{equation} 
\hfill \break
One can describe the propagation of a Gaussian beam by transforming this complex source by using \cite{siegman1986lasers}:

\begin{equation} \label{eq:q1Toq2}
q_2 = \frac{A q_1 + B}{C q_1 + D}
\end{equation} 
\hfill \break
where the subscripts 1 and 2 denote the initial and final source position and variables $A$ to $D$ are entries in $M = [A~B; C~D]$; an optics ray tracing propagation transfer matrix. In the matrix ; denotes the next lower row. \\
\\
The intensity of the electric field $I= \varepsilon_0 c |E|^2 /2$, for Eq.~\ref{eq:laserEfield}, is given as:

\begin{equation}\label{eq:laserIntensity}
    I(x,y,z)=I_0\frac{w_0^2}{w^2(z)}\exp\left(-2\frac{x^2+y^2}{w(z)^2}\right)
\end{equation}
\hfill \break
where $\varepsilon_0$ is the vacuum permittivity and is included in $I_0$. Note that the half width of the beam $w$ is measured from where the laser is at $e^{-2}$ of its peak intensity to its center line at some $z$-value. Note that the intensity half width at $e^{-2}$ of its peak is equal to the $e^{-1}$ half width for the electric field, due to the intensity being obtained by squaring the electric field.

A 3D model for a laser beam that uses rays projected from the $xy$-plane to represent the intensity of the beam, 
is described in \cite{pang2021modeling}. The equivalent spatial distribution is given by Eq.~\ref{eq:laserIntensity} with  $I_0=1$ \cite{pang2021modeling} and 
the ray origins given by $I(x,y,0)$. Note that Eq.~\ref{eq:laserIntensity} has the same form as Eq.~\ref{eq:samsol} which is used to model DED-PF powder streams. The spectrum of the rays given in Eq.~13 of \cite{pang2021modeling} based on the angles $(\theta_x,\theta_y)$ or equivalently based on $(k_x,k_y)=k(\tan(\theta_x),\tan(\theta_y))$ can be written as:
\begin{align} \label{eq:laserAngularDistPang}
P_2(k_x,k_y)&=\frac{w_0^2}{2 \pi} \exp\left(-2\frac{k_x^2 + k_y^2}{s_0^2}\right)\\
\Leftrightarrow P_1(\theta_x,\theta_y)&=\frac{w_0^2k^2}{2 \pi\cos(\theta_x)\cos(\theta_y)} \exp\left(-2k^2\frac{\tan^2(\theta_x) + \tan^2(\theta_y)}{k_0^2}\right)\label{eq:angular_spectrum}
\end{align} 
\hfill \break
with $s_0 = w_0/z_R$ which approximates $\theta_0$ in \cite{crooker2006representation} at small angles, $k_x=k\tan(\theta_x)$ and $k_0= 2 / w_0$. 

Note that the ray spatial and tangential component distributions can be used directly as an initial condition for the particles in a Lagrangian simulation \cite{martinezmarchese2022application}. The significance of speed variation as well as the drag force on the particle due to the surrounding medium using this new model will be shown later via Lagrangian simulations in Section \ref{sec:lagrangianParticleSimulation}.  
\newpage
\section{Theoretical and experimental methods}

\subsection{Particle tracking and trajectory extrapolation} \label{subsec:particleTracking}
In order to determine the behaviour of the particles in the powder stream, the python particle tracking code Trackpy \cite{allandanielb2021} was used to extract path information from high-speed video. The video was obtained by filming the powder stream produced by manually feeding SS 316L powder to a gravity hopper connected to a nozzle. The nozzle disperses the powder with a center-body before moving to a converging nozzle with an exit diameter of 0.9 mm, in order to produce the Gaussian distribution with a linearly increasing IRW as observed in the DED-PF literature~\cite{martinezmarchese2022application}. The SS 316L powder used is from North American H\"{o}gan\"{a}s (316 L-5320), item \# 111903. The mass average diameter was determined to be \SI{89}{\micro m}, measured using a Camsizer X2 from Retsch GmbH.

Trackpy uses the Crocker-Grier centroid-finding algorithm to find particles in a frame \cite{crocker1996methods}. The settings used for finding the particle positions were the following: The frames were cropped to not include the top of the ultrasound array used to produce the force field \cite{martinezmarchese2022application}, an estimated particle size of 7 pixels (found by taking the square root of the avg. pixel area for particles found in \cite{martinezmarchese2022application}), a minimum `mass' (brightness) value of 200 was used. The setting used for extracting paths were the following: A search range of 4 pixels/frame (approximate speed of particles) and a memory of 1 frame (number of frames particle does not have to be present in a frame) were used. The generated paths were filtered by not including short paths (less than 40 frames), and only including paths with particles with a mass variable (Trackpy dimensionless variable) between 500 and 2000, a size between 1 and 1.5 pixels and a circular eccentricity of less than 0.3.

Afterwards the paths are centered in $x$, rotated, shifted in $z$ such that the nozzle position is at $z=0$ and scaled by using the pixel size of \SI{72.9}{\micro m} measured in \cite{martinezmarchese2022application}. The rotation angle is calculated such that it undoes the offset between the $x$-averages of the Gaussian fits of the powder stream at the top and bottom of the sum of frames image \cite{martinezmarchese2022application}, which can happen due to camera misalignment. The paths are also filtered by checking that subsequent line segments formed by two subsequent particle positions have an angle of less than 3\textdegree \ (using the absolute value of their dot product) and do not change direction.

The paths are then fitted to a particle trajectory analytic model \cite{siklos}. It is the analytic solution for a particle's 2D trajectory taking into account gravitational acceleration and the drag force on the particle from the surrounding medium (still air) assuming a linear relationship with the particle's relative speed (Stokes' drag law). Its applicability was checked by computing the maximum Reynolds number ($Re$) for all feasible trajectories as described later in this section. For this drag law to be applicable, $Re \sim 1$. With these trajectory fits, one can calculate accurate particle statistics such as the angle of the path with respect to the $x$-direction at any $z$-value and extrapolate the particle paths up to the nozzle position, which is obstructed by the ultrasound array producing the force field for the data from \cite{martinezmarchese2022application} used in this article. 

If the diagonal of the bounding box of the path to be fitted has an angle of 0.29\textdegree, corresponding to an aspect ratio of $5.0 \times 10^{-3}$, a straight line fit is carried out. This is because the fitting problem is ill posed; when the paths are too close to a vertical line, one can have a large range of solutions for the initial speed. For the fit and evaluation of the line, the axes are switched to prevent numerical errors due to high slope values.

To define a path, the following variables are required: the initial $x_0$ position, the initial velocity components $u_{0x}$ and $u_{0z}$, and the drag factor dependent on the particle size $k$. It is assumed the particles all start at the nozzle $z$-location ($z=0$). One can obtain these variables by minimizing the following:

\begin{equation} \label{eq:pathModel1}
f = \sum_{i=1}^N (h_i - z_i)^2
\end{equation} 
\hfill \break
where $N$ is the number of points in a path, $h_i$ is the modeled $z$-position for each point and $z_i$ is the actual $z$-position for each point. The model $z$-positions can be found from the actual $x$-positions $x_i$ by rearranging the equation for $x$ in \cite{siklos} to get $t$: 

\begin{equation} \label{eq:pathModel2}
t_i = - \frac{1}{k} \log\left(1 - \frac{k}{u_{0x}} (x_i - x_0)\right)
\end{equation} 
\hfill \break
where $k$ is Stokes's law constant divided by the particle mass \cite{siklos, ahmadi}:

\begin{equation} \label{eq:kNorm}
k = \frac{3 \pi \mu d_p}{m_p}
\end{equation}
\hfill \break
where $\mu$ is the air viscosity, $d_p$ is the particle diameter and $m_p$ is the particle mass. Afterwards the $t_i$ values are used to find $h_i$ using the equation for $z$ in \cite{siklos}:

\begin{equation} \label{eq:pathModel3}
h_i = t_i \frac{g}{k} + \frac{1}{k} \left(u_{0z} - \frac{g}{k}\right) \left(1 - e^{-kt_i}\right)
\end{equation} 
\hfill \break
where $g$ is the gravitational acceleration. Note one can do the same minimization considering the $x$-locations, however Eq.~\ref{eq:pathModel3} cannot be rearranged analytically for $t$. One can minimize $f$ by optimizing the path variables for each path using the function \verb|fmincon()| in MATLAB, using the `Trust region reflexive' algorithm \cite{coleman1996interior} with a function tolerance of $1.0 \times 10^{-8}$. It was found that providing the gradient of $f$ with respect to the path variables to \verb|fmincon()| produces a more accurate fit. This gradient can be calculated using the multivariable chain rule as follows:

\begin{equation} \label{eq:pathModelGrad1}
\begin{bmatrix}
\boldsymbol{\rm Q}_{x_0} \\
\boldsymbol{\rm Q}_{u_{0x}} \\
\boldsymbol{\rm Q}_{u_{0z}} \\
\boldsymbol{\rm Q}_k
\end{bmatrix}
=
\begin{bmatrix}
(\boldsymbol{\rm h}_t \circ \boldsymbol{\rm t}_{x_0}) \boldsymbol{\rm f}_h^T \\
(\boldsymbol{\rm h}_t \circ \boldsymbol{\rm t}_{u_{0x}}) \boldsymbol{\rm f}_h^T \\
\boldsymbol{\rm h}_{u_{0z}} \boldsymbol{\rm f}_h^T \\
\boldsymbol{\rm h}_k \boldsymbol{\rm f}_h^T
\end{bmatrix}
\end{equation}
\hfill \break
where the variables in bold are 1 by $N$ matrices, the subscript stands for partial derivation with respect to the subscript and $\circ$ stands for element-wise multiplication. The matrix $\boldsymbol{\rm f}_h$ can be found as follows:

\begin{equation} \label{eq:pathModelGrad2}
\boldsymbol{\rm f}_h = 2(\boldsymbol{\rm h} - \boldsymbol{\rm z})
\end{equation}
\hfill \break
Each entry of the other matrix derivatives are the element-wise derivatives of Eqs.~\ref{eq:pathModel2} or \ref{eq:pathModel3}. The optimization function was used with the following bounds:

\begin{align}
-0.02 < \ &x_0 < 0.02 \\
-0.5 < \ &u_{0x} < 0.5 \\
0.0 < \ &u_{0z} < 0.5 \\
1.0406 < \ &k < 104.0625
\label{secSupp:v0sol}
\end{align}
\hfill \break
The bounds enforce for tracks that start close to the nozzle, and that they do not have a high starting velocity or positive $z$-velocity component. The bounds in Eq.~\ref{secSupp:v0sol} were calculated using Eq.~\ref{eq:kNorm} and particle sizes of 20 and \SI{200}{\micro m}. The fits for 2168 paths of various lengths takes about 20 seconds in MATLAB.  

The resulting paths were filtered as follows: paths with an initial speed ratio ($u_{0x}/u_{0z}$) of 5 or more or with an initial position $x_0$ further away than 2.5 mm from the nozzle were not used. The final number of paths used for powder stream statistics is 2043.

To check the low $Re$ number assumption for the use Stokes' law, for each path fitted with the linear drag model, the travel time of the particle from $z=0$ to the maximum measured $z$ value (72 mm) from TrackPy was found by numerically solving Eq.~\ref{eq:pathModel3} for $t_i$ and using 72 mm for $h_i$ using the function \verb|fzero()| in MATLAB with a 0 to 1s bound. Afterwards this time is used to obtain the velocity components at the end of the path. The velocity components are given by \cite{siklos} 

\begin{align} 
u_{x} &= u_{0x} e^{-kt_i} \label{eq:pathModelUx} \\
u_{z} &= \frac{g}{k} + \left(u_{0z} - \frac{g}{k}\right) e^{-kt_i}. \label{eq:pathModelUz}
\end{align} 
\hfill \break
The maximum $Re$ number for all tested paths using the velocity magnitude and air properties was 10. However, only 19 of the 1772 trajectories fitted with the linear drag model (the rest being the paths fitted with a line since they are close to vertical) had a value greater than 6.

\subsection{Determination of ray statistics in terms of divergence angle and nozzle width}
The $e^{-1}$ IRW and divergence angle measured in \cite{martinezmarchese2022application} (and scaled to values considering $e^{-2}$ IRW) corresponds to the half width $w_m$ and $tan(\theta_m) = dw/dz(w_m)$. Using these equations and Eq.~\ref{eq:w}, one can solve for $w_0$ and $z_R$ using Eqs.~\ref{eq:w0accurate} and \ref{eq:z0accurate}.

\begin{align}  
w_0 &= \sqrt{w_m^2 - \tan(\theta_m) w_m z_m} \label{eq:w0accurate} \\
z_R &= w_0 \sqrt{\frac{z_m}{w_m} tan(\theta_m)} \label{eq:z0accurate}
\end{align} 
\hfill \break
The divergence angle $\theta_0$ equals $\arctan(w_0/z_R)$ since $dw/dz$ approaches $w_0/z_R$ as $z \to \infty$. The powder behaviour observed in \cite{martinezmarchese2022application} suggested that the there is a linear function between speed and position normal to the powder stream axis at some distance from the nozzle exit. One would expect the slope of this line to become steeper as one gets close to the nozzle, as well as the coefficient of determination $r^2$ to approach 1 far away from the nozzle and approach zero when close to the nozzle. 
The expected value of $r^2$ is given by \cite{chatterjee2015regression}:

\begin{equation} \label{eq:R2}
\mathbb{E}\left(r^2\right) = 1 - \mathbb{E} \bigg ( \frac{SS_{res}}{SS_{T}} \bigg ) = 1 - \frac{\mathbb{E}(SS_{res})}{\mathbb{E}(SS_{T})}
\end{equation} 
\hfill \break
where $SS_{res}$ is the sum of squares of the sampled residuals (difference between predicted and measured data) and $SS_{T}$ is the sum of squares of the sampled data. The assumption made for the last term is that samples from $SS_{res}$ can be considered independent of samples from $SS_{T}$. 

From Eq.~1 in \cite{pang2021modeling} and Eq.~\ref{eq:angular_spectrum} one can do a similar calculation as Eqs.~28-32 in \cite{pang2021modeling} and derive the probability density for $\theta_x$ given $x$ with $k_1(z)= k / \sqrt{z^2k_0^2+w_0^2k^2}$:

\begin{equation}
    f_{\theta_x}(\theta_x|x)=\frac{1}{\sqrt{2\pi}k_1(z)\cos^2(\theta_x)}\exp\left(-k^2\frac{\left(\tan(\theta_x)-\frac{x}{R(z)}\right)^2}{2k_1^2(z)}\right)\label{eq:condDensityAngleXFreeProp}
\end{equation}
\hfill \break
One can see from Eq.~\ref{eq:laserIntensity} that the probability density of $x$ is:

\begin{equation}
    f_x(x)=\sqrt{\frac{2}{\pi w^2(z)}}\exp\left(-2\frac{x^2}{w^2(z)}\right)\label{eq:densityXFreeProp}
\end{equation}
\hfill \break
With Eqs.~\ref{eq:condDensityAngleXFreeProp}-\ref{eq:densityXFreeProp}, one can calculate the expected value of a random variable $a$ using Eqs \ref{eq:Ethetax}-\ref{eq:Efora}.

\begin{align}
    \mathbb{E}_{\theta_x}(a) &= \int_{-\frac{\pi}{2}}^{\frac{\pi}{2}}a\cdot f_{\theta_x}(\theta_x|x)\,d\theta_x \label{eq:Ethetax}\\
    \mathbb{E}_x(a) &= \int_{-\infty}^{\infty}a\cdot f_x(x)\,dx \label{eq:Ex}\\
    \mathbb{E}(a) &= \mathbb{E}_{x}\left(\mathbb{E}_{\theta_x}(a)\right) \label{eq:Efora}
\end{align}
\hfill \break
To calculate Eq.~\ref{eq:R2}, one needs:

\begin{equation}\label{eq:ESSres}
    \mathbb{E}(SS_{res})=\mathbb{E}\left(\sum_{n=1}^N \left(y_n - \mathbb{E}_{\theta_x}(y_n)\right)^2\right)
\end{equation}
and:

\begin{equation}\label{eq:ESST}
    \mathbb{E}(SS_T)=\mathbb{E}\left(\sum_{n=1}^N \left(y_n - \mathbb{E}(y_n)\right)^2\right)
\end{equation}
\hfill \break
with $y_n$ being the measurements of the random variable $Y:=\theta_x|x$ and the number of measurements $N\rightarrow\infty$. The exact calculation of these two expected values are given in the appendix as a power series. To simplify the calculations, we present here the result in paraxial approximation, i.e., $\tan(\theta_x)\approx\theta_x$ corresponds to using the first term of the power series. Under this approximation:

\begin{align*}
    &\mathbb{E}(SS_{res})=N\frac{k_1^2(z)}{k^2}\\
    &\mathbb{E}(SS_T)=(N-1)\left(\frac{w^2(z)}{4R^2(z)}+\frac{k_1^2(z)}{k^2}\right)
\end{align*}
Therefore, for $N\rightarrow\infty$:

\begin{align}
    \mathbb{E}\left(r^2\right) =1-\frac{\frac{k_1^2(z)}{k^2}}{\frac{w^2(z)}{4R^2(z)}+\frac{k_1^2(z)}{k^2}}=1-\frac{1}{1+\frac{k^2w^2(z)}{4R^2(z)k_1^2(z)}}\label{eq:r2}
\end{align}
\hfill \break
By defining $\theta_1(z)=\theta_0/\sqrt{1+z^2/z_R^2}$ and using the paraxial approximation a second time for $k_1^2/k^2\approx \theta_1^2/4$ this can be approximated by:
\begin{equation}\label{eq:R2full}
    \mathbb{E}\left(r^2\right) =1-\frac{\theta_1^2(z)}{\theta_1^2(z)+\frac{w^2(z)}{R^2(z)}}=1-\frac{1}{1+\frac{w^2(z)}{R^2(z)\theta_1^2(z)}}
\end{equation}
\hfill \break
It follows that $\mathbb{E}\left(r^2\right)\rightarrow1$ for $z\rightarrow\infty$ and $\mathbb{E}\left(r^2\right)\rightarrow0$ for $z\rightarrow0$ in paraxial approximation.

Eq.~\ref{eq:R2full} explains the close to linear behaviour between sideways particle speed and distance from the nozzle center-line when far away from the nozzle as observed in \cite{martinezmarchese2022application} and spatial independence of this speed as expected from the ray representation of a Gaussian beam at the nozzle area.

\subsection{Lagrangian particle simulation in the presence of a force field}
The initial particle speed distribution for the Gaussian beam ray representation model can also be used as an initial condition in the 3D Lagrangian particle tracking model described in \cite{martinezmarchese2022application}. This model can be used to calculate the sound pressure and resulting force field on the particles. By calculating the force field at a required particle location, one can perform Lagrangian particle tracking by computing particle paths from an initial to a final position at a required time interval. This can be done using a 4\textsuperscript{th} order Runge-Kutta (RK4) method \cite{faires2012numerical} with a given fixed time step. The library is described in more detail in the supplementary information of \cite{martinezmarchese2022application}. 

The initial particle concentration at the nozzle position can be approximated by a group of particles, chosen such that their probability function approaches the particle concentration function. A Monte Carlo (MC) simulation \cite{kalos2009monte} can be run with enough particles to have a good estimate of the particle concentration. The model uses a speed standard deviation $s_{STD}$ of 0.032 m/s as measured in \cite{martinezmarchese2022application}, a gas particle drag force model and does not consider gravitational acceleration.

The initial particle spatial and tangent component distributions are generated using Eqs.~\ref{eq:laserIntensity} and \ref{eq:laserAngularDistPang} as shown in Algorithm \ref{alg:initialConditions}. The algorithm uses the Box-Muller algorithm to produce one or two normally distributed values from $X_n$, sampled from a 0 to 1 uniform distribution \cite{kalos2009monte}. A unit vector is then computed from the randomly sampled tangential components and then multiplied by the sampled speed $U$.

\begin{algorithm}
\caption{Generation of initial positions and speeds for Lagrangian simulation } \label{alg:initialConditions}
\begin{algorithmic}
\State $N \gets 1$
\While{$N \leq n_r$}
\State $r \gets (w_0/2) \sqrt{log(-2 X_1)}$ \Comment{Two variable Box-Muller for particle position}
\State $x(N),y(N) \gets r \ cos(2 \pi X_2), r \ sin(2 \pi X_2)$
\\
\State $t \gets (s_0/2) \sqrt{log(-2 X_3)}$ \Comment{Two variable Box-Muller for trajectory tangential components}
\State $tan(\theta_x), tan(\theta_y) \gets t \ cos(2 \pi X_4), t \ sin(2 \pi X_4)$
\\
\State $x(N),y(N) \gets x(N) + d_p tan(\theta_x), y(N) + d_p tan(\theta_y)$ \Comment{Calculate positions upstream of force field}
\State $z(N) \gets 0.0$ \Comment{10 mm upstream of sound focus point}
\\
\State $U \gets u_{avg} + u_{STD} \sqrt{log(-2 X_5)} \ cos(2 \pi X_6)$ \Comment{One variable Box-Muller to include measured speed variation}
\\
\State $|u| = \sqrt{1 + tan^2(\theta_x) + tan^2(\theta_y)}$
\State $u_{0x}(N) = U tan(\theta_x)/|u|$ \Comment{Components divided by $U$ correspond to the ray's direction cosines \cite{colbourne2019representation}}
\State $u_y(N) = U tan(\theta_y)/|u|$
\State $u_{0z}(N) = U /|u|$
\\
\State $N = N+1$
\EndWhile
\end{algorithmic}
\end{algorithm}

Note that the $w_0$ and $s_0$ values are divided by two to match the variance given by Eqs.~\ref{eq:laserIntensity} and \ref{eq:laserAngularDistPang} when using the Box-Muller method which gives values for a standard normal distribution \cite{kalos2009monte}. The particles are projected forward a distance of $d_p$ = 44.2 mm - 20 mm = 22.2 mm in a straight line to before 10 mm of the sound focus point, upstream of the sound interaction region \cite{martinezmarchese2022application}, to produce a close match with the measured PCD with no applied force field. Running the Lagrangian simulation for longer distances can produce a mismatch between the measured and calculated PCDs with no applied force field due to air drag.

The Lagrangian simulation was run by simulating the force field due to a vortex sound field \cite{marzo2015holographic}. The sound field was produced by an array of ultrasound transducers running at different applied voltages. At \SI{0}{V} there is no applied force field, and at 10 and \SI{16}{V} the force fields produced had maximum absolute values of $2.92 \times 10^{-9}$ and $7.47 \times 10^{-9}$ N per particle \cite{martinezmarchese2022application}.

The simulation results in Section \ref{sec:lagrangianParticleSimulation} were run using 2 time steps for the runs at 0 V and 4 time steps for all other voltages. The number of paths calculated were 12.7e4, 8.1e4 and 3.6e4 for the 0, 10 and 16 V simulations to achieve the required accuracy. The simulations were done using a $\Delta r$ (radial sampling interval) value of 0.5 mm to compute the concentration function. The computing times for the simulations were 60, 72 and 34 minutes for the 0, 10 and 16 V simulations. A convergence study was used to show the above simulation parameters produce a converged solution, with an estimated error of less than 2\% for the RK4 particle tracking and less than 4\% for the MC simulation. All simulations in this report were carried out using a Dell OptiPlex desktop, with an Intel Core i7 3.6 GHz CPU processor and 16 GB of RAM. More information on the convergence study procedure is given in the supplementary information of \cite{martinezmarchese2022application}.

\subsection{The focusing sound field as an optical transfer matrix} \label{sec:abcd}
As compared to the Fresnel-Kirchhoff integral \cite{siegman1986lasers}, the Gaussian beam ray representation considers rays without a phase component. This representation produces an accurate intensity distribution of the beam if the rays do not change direction to a large degree, such as in situations of high levels of refraction or in diffraction, where closely spaced rays would start to constructively or destructively interfere due to phase differences. Therefore the Gaussian beam ray representation is an accurate model of a Gaussian beam propagating in free space, or when considering propagation in an optical system where the paraxial approximation is adequate \cite{crooker2006representation}. This means that after adequately modeling the powder stream with the Gaussian beam ray representation, one could use optical transfer matrix methods for Gaussian beams \cite{siegman1986lasers} to model the motion of the powder stream when subjected to a force field as long as the particle deflection is not large.

\newpage
For example, one can derive the particle motion in a radially symmetric force field where the force points towards the $z$ axis and decreases linearly with distance to the $z$ axis (close to $l_{offset}$) as shown in Fig.~\ref{fig:vortexFieldProps}. The resulting equations of motion are the following \cite{martinezmarchese2022application}:

{\begin{align}
r_f &= r_i cos(\eta l_i) + \frac{tan(\theta_i)}{\eta} sin(\eta l_i) \label{eq:motionLinerForceField1} \\
\theta_f &= tan^{-1}[-r_i \eta sin(\eta l_i) + tan(\theta_i) cos(\eta l_i)] \label{eq:motionLinerForceField2}
\end{align}}
\hfill \break
where $r_i$ and $r_f$ are the initial and final particle normal to the $z$ axis and $\theta_i$ and $\theta_f$ are the initial and final angles with respect to the $z$ axis. $\eta = \sqrt{F_a/(l_{offset} m_p s_{pz}^2)}$, where $F_a$ is a quarter of the peak force magnitude and $s_{pz}$ is the average downward component of the particle velocities. The parameter $l_i$ is the length along which the sound field is considered to affect the particles and $l_{offset}$ is half the length from the powder stream center-line to the location of peak force magnitude, this is the length where the force field might be considered to be linear \cite{martinezmarchese2022application}. The last two parameters described are shown in Fig.~\ref{fig:vortexFieldProps}.

Using Eqs.~\ref{eq:motionLinerForceField1} and \ref{eq:motionLinerForceField2}, and considering small particle deflections ($tan(\theta) \approx \theta$), a symmetric sound radiation force field close to the axis of symmetry, due to a vortex sound field, can be modeled as an optical transfer matrix given by:

\begin{equation} \label{eq:ABCDgrin}
M_{GRIN}(l_i) = \begin{pmatrix} A & B \\ C & D \end{pmatrix}  = \begin{pmatrix} cos(\eta l_i) & \frac{1}{\eta} sin(\eta l_i) \\ -\eta sin(\eta l_i) & cos(\eta l_i) \end{pmatrix}
\end{equation}
\hfill \break
This transfer matrix corresponds to a dielectric rod with a radially varying optical index profile given by Eq.~\ref{eq:nProfile}, called a graded refractive index (GRIN) lens \cite{haus1984waves}.

\begin{equation} \label{eq:nProfile}
n = n_0 \bigg ( 1 - \frac{1}{2} \alpha^2 r^2 \bigg )
\end{equation}
\\
With a lens center-line index of diffraction $n_0 = 1$ and a parabolic dependence of $\alpha = 1/\eta$. Note the determinant of Eq.~\ref{eq:ABCDgrin} is one. This means that since we can model the powder stream with representing a Gaussian \cite{crooker2006representation}, we can model DED-PF powder stream focusing as a Gaussian beam being calculated using a 2x2 transfer matrix, so called ABCD system. The pitch of the GRIN lens is defined as $\eta l_i/(2\pi)$. A pitch of $1/4$ means that a point source on one side of the lens becomes a parallel ray to the z-axis on the other side \cite{van2003foundations}.

The transfer matrix can be used to plot the PCD along $z$. Using $w_0$ and $z_R$ by solving Eqs.~\ref{eq:w0accurate} and \ref{eq:z0accurate}, one can compute an effective $\lambda$ for the powder stream, using Eq.~\ref{eq:zR}.

The start of the laser propagation is at its waist, where $R$ is considered zero, with $q$ given by $j z_R$ (solving Eq.~\ref{eq:complexQ}) \cite{jung2010numerical}. Then one can apply Eq.~\ref{eq:q1Toq2} repeatedly using ABCD = [1 $l$; 0 1], which is the transfer matrix for free space propagation. This is computed at different equally spaced distances $l$ along $z$, with the start of the beam considered at $z=0$. When one reaches what is considered the upstream position of the force field, one can keep using Eq.~\ref{eq:q1Toq2}, but using the following transfer matrix:

\begin{equation} \label{eq:totalM1}
M_1(z) = M_{GRIN}(z - l_{fs}) M_{fs}(l_{fs})
\end{equation}
\\
where $M_{fs}(l_{fs})$ stands for the free space transfer matrix and $l_{fs}$ is the distance from the start of the beam to the force field. Note that one can combine transfer matrices, by multiplying them, with the matrices closest to the source of the beam on the right \cite{siegman1986lasers}. One can apply this again downstream of the force field:

\begin{equation} \label{eq:totalM2}
M_2(z) = M_{fs}(z - l_{fs} - l_i) M_{GRIN}(l_i) M_{fs}(l_{fs})
\end{equation}
\\
to find $q$ values at any position downstream of the force field.

At each $z$-position, one can compute the half width by solving for $w$ using Eq.~\ref{eq:complexQ}:

\begin{equation} \label{eq:wSol}
w = \sqrt{-\frac{\lambda}{\Im(1/q) \pi}}
\end{equation}
\\
And then use the normalized (such that its polar integral is equal to 1) version of Eq.~\ref{eq:laserIntensity} with $y=0$, times Eq.~\ref{eq:np}:

\begin{equation} \label{eq:normalizedI}
I = \frac{2 n_p}{w^2 \pi} \exp\left(-2 \frac{x^2}{w^2}\right)
\end{equation}
\\
One can then use Eq.~\ref{eq:normalizedI} to compute the PCD along $x$. For a $z$-range of 0 to 140 mm, and an $x$-range of -15 to 15~mm, both with an interval of 0.1 mm, the above calculations take about 0.2 seconds in MATLAB. 

\newpage
Using the $q$ value right after the force field, using $M_2(l_{fs} + l_i)$, one can compute the distance from the force field to the waist of the focused beam spot using \cite{jung2010numerical}:

\begin{equation} \label{eq:zs}
z_s = -\Re(q)
\end{equation}
\\
This is derived by writing the location of the focus $q_f$ as $q + z$ and using the fact that at a focus point the radius $R$ is infinite, meaning $q_f$ should be imaginary, so $\Re(q) + z = 0$ \cite{mittlemanSlides}.

The half width at the focus is given by \cite{jung2010numerical}:

\begin{equation} \label{eq:ws}
w_s = \sqrt{\frac{\lambda \Im(q)}{\pi}}
\end{equation}
\\
This is derived by noting that $q_f$ is equal to $-j \Im(q)$ from the result for $z_s$, and solving for $w$ using Eq.~\ref{eq:complexQ} \cite{mittlemanSlides}. An explicit expression for Eqs.~\ref{eq:zs} and \ref{eq:ws} is given in \cite{jung2010numerical}. Eqs.~\ref{eq:zs} and \ref{eq:ws} are valid only for $z_s \geq 0$, i.e.; the beam waist is at the boundary of or after the GRIN lens.

\subsection{Optimal powder focusing parameters based on optical analog model} 
\begin{figure}[!ht]
\center
\includegraphics[width=17cm]{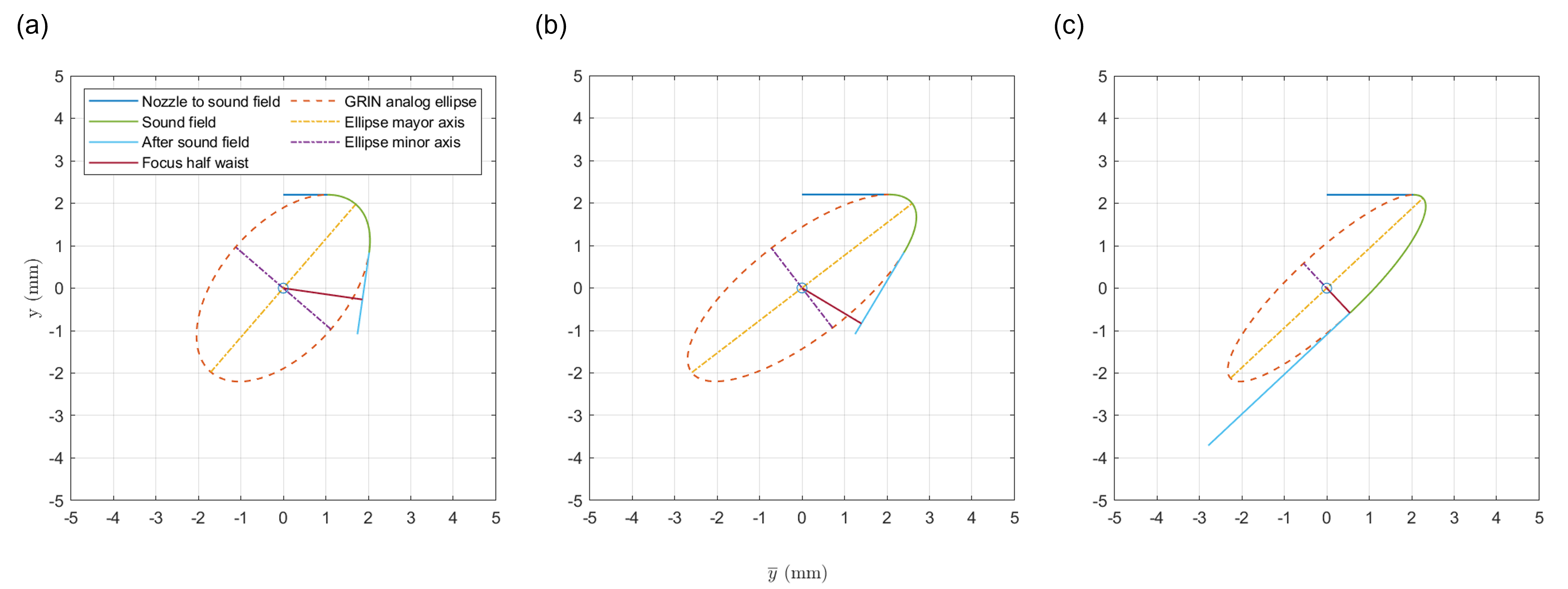}\\ 
\caption{$y$\textendash$\overline{y}$ diagram corresponding to powder exiting a nozzle and interacting with a focusing sound radiation force field. (a) First case; $w_0$ = 2.2 mm, $z_R$ = 21.2 mm, $l_i$ = 20.0 mm, $l_{fs}$ = 10.0 mm and $\eta$ = 59.4 m\textsuperscript{-1}. (b) Case shown in a with higher nozzle to sound field distance ($l_{fs}$ = 19.7 mm). (c) Case shown in b with higher applied transducer array voltage ($\eta$ = 92.0 m\textsuperscript{-1})}
\vspace{0.5cm}
\label{fig:yyBarDiags}
\end{figure}

Eqs.~\ref{eq:zs} and \ref{eq:ws} can be written explicitly and are given by \cite{jung2010numerical}:

\begin{align} 
z_s &= \frac{[1 + (l_{fs}/z_R)^2 - 1/(z_R \eta)^2]sin(2 \eta l_i) - 2 l_{fs}/(z_R \eta) cos(2 \eta l_i)}{2 z_R \eta \sqrt{sin^2(\eta l_i) +  [cos(\eta l_i) - \eta l_{fs} sin(\eta l_i)]^2/(z_R \eta)^2}} \label{eq:zsAnalytic} \\
w_s &= \frac{w_0}{z_R \eta \sqrt{sin^2(\eta l_i) +  [cos(\eta l_i) - \eta l_{fs} sin(\eta l_i)]^2/(z_R \eta)^2}} \label{eq:wsAnalytic}
\end{align}
\\
where $w_s$ corresponds to the minimum powder spot size at a distance $z_s$ from the sound force field. The propagation of the beam along an optical system can also be described graphically, using a $y$\textendash$\overline{y}$ diagram. A beam can be described by two rays where the rays should satisfy a Lagrange invariant \cite{kessler1992yy}. The coordinates of the points in the lines and curves on the $y$\textendash$\overline{y}$ diagram are the heights of these two rays, parametrized by the $z$-distance along the optical system.  An example of the diagram for the analog optical setup in this article is shown in Fig.~\ref{fig:yyBarDiags} (a). In this case the two rays used are the divergence ray (height $\overline{y}=0$ and angle $\overline{u} = \theta_0$) and the waist ray (height $y=w_0$ and angle $u=0$) \cite{kessler1992yy}. The diagram can be calculated by applying the same transfer matrices in Section \ref{sec:abcd} for the two rays being considered. The $y$\textendash$\overline{y}$ curve for a GRIN lens can be shown to correspond to a rotated ellipse \cite{rogers1988y, harrigan1988use}. Its dimensions are also derived in \cite{kessler1992yy}.

Assuming that reducing the width of the powder nozzle ($\approx w_0$) will proportionally decrease the divergence angle \cite{vilar1999laser} ($z_R$ is approximately constant with respect to $w_0$), one can find the optimal $w_0$ and $\theta_0$ values that minimize $w_s$. Note that Eq.~\ref{eq:zR} cannot be used directly since there is no physical analog to the light wavelength $\lambda$ in this model. From inspection of Eq.~\ref{eq:wsAnalytic} one can see that smaller $w_0$ values produces a smaller $w_s$ value. Expanding the denominator of Eq.~\ref{eq:wsAnalytic} one can also note that a larger $z_R$ value gives a lower $w_s$ value which corresponds to minimizing $\theta_0$ for a fixed $w_0$.

\begin{figure}[!ht]
\center
\includegraphics[width=12cm]{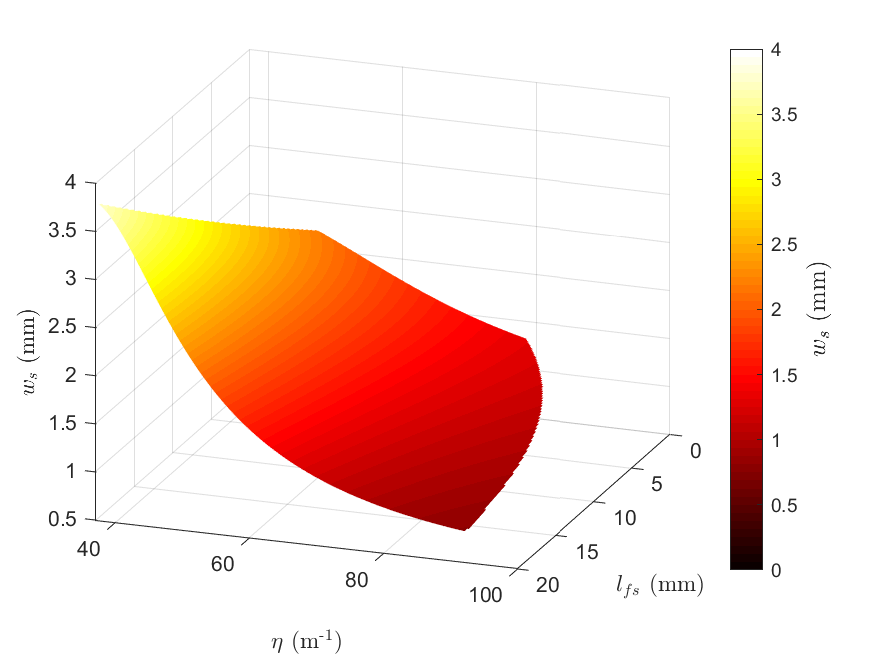}\\ 
\caption{Powder waist half width downstream of sound field as a function of $l_{fs}$ and $\eta$.}
\vspace{0.5cm}
\label{fig:wsSurfOpt}
\end{figure}

A plot of Eq.~\ref{eq:wsAnalytic} for different $l_{fs}$ and $\eta$ values for cases where $z_s \geq 0$ is shown in Fig.~\ref{fig:wsSurfOpt}. One can see a weak inverse relationship between $l_{fs}$ and $w_s$ at high $\eta$ values.

This can also be observed in Fig.~\ref{fig:yyBarDiags} (b), where the $y$\textendash$\overline{y}$ diagram is plotted for a larger $l_{fs}$ value, which reduces the minor axis length of the ellipse. Note that for larger $\eta$ values $w_s$ is reduced, however the powder spot size should remain outside or at the edge of the sound field. As observed in Fig.~\ref{fig:yyBarDiags} (c), increasing $\eta$ moves the start of the $y$\textendash$\overline{y}$ diagram line after the GRIN lens ellipse toward the minor axis. One can find the beam waist of this beam from its closest (normal) distance from the line in the diagram to the diagram's origin, which corresponds to when this line exits the GRIN lens ellipse at its minor axis. This also corresponds to the beam's waist location where the beam focuses. Therefore, one should use the highest $\eta$ (proportional to the applied ultrasound transducer voltage) possible as long as the focus point is outside or at the boundary of the sound field ($\eta$ in Eq.~\ref{eq:zsAnalytic} such that $z_s = 0$, shown in Fig.~\ref{fig:yyBarDiags} (c)).

Taking into account material utilization, there is also a constraint on how long $l_{fs}$ can be; the width of the beam at the start of the force field should be smaller than what is considered twice the force field's `radius'. This is the point where the force field is not considered to be linear in the radial direction \cite{martinezmarchese2022application}. The required distance is given by using Eq.~\ref{eq:w} and solving for $z$ using $z=l_{fs}$:

\begin{equation} \label{eq:lfs}
l_{fs} = z_R \sqrt{\left(\frac{kr}{w_0}\right)^2 - 1}
\end{equation}
\\
where $r$ is the effective force field radius and $k$ is a factor smaller than one to take into account `limiting rays' \cite{harrigan1984some}, corresponding to the fact that the beam for some distance into the force field is still expanding. Therefore, to find the optimal $l_{fs}$ and $\eta$ values one should use Eq.~\ref{eq:lfs} to find $l_{fs}$ and then use this value to find $\eta$ such that the numerator of Eq.~\ref{eq:zsAnalytic} is equal to zero.

A more detailed analysis taking this into limiting rays requires $\eta$ to be known \cite{harrigan1984some}. This could be done finding the optimal $l_{fs}$ and $\eta$ values iteratively.

\section{Results and discussion}

\subsection{Particle path statistical characteristics}

\begin{figure}[!ht]
\center
\vspace{0.2cm} 
\includegraphics[width=14cm]{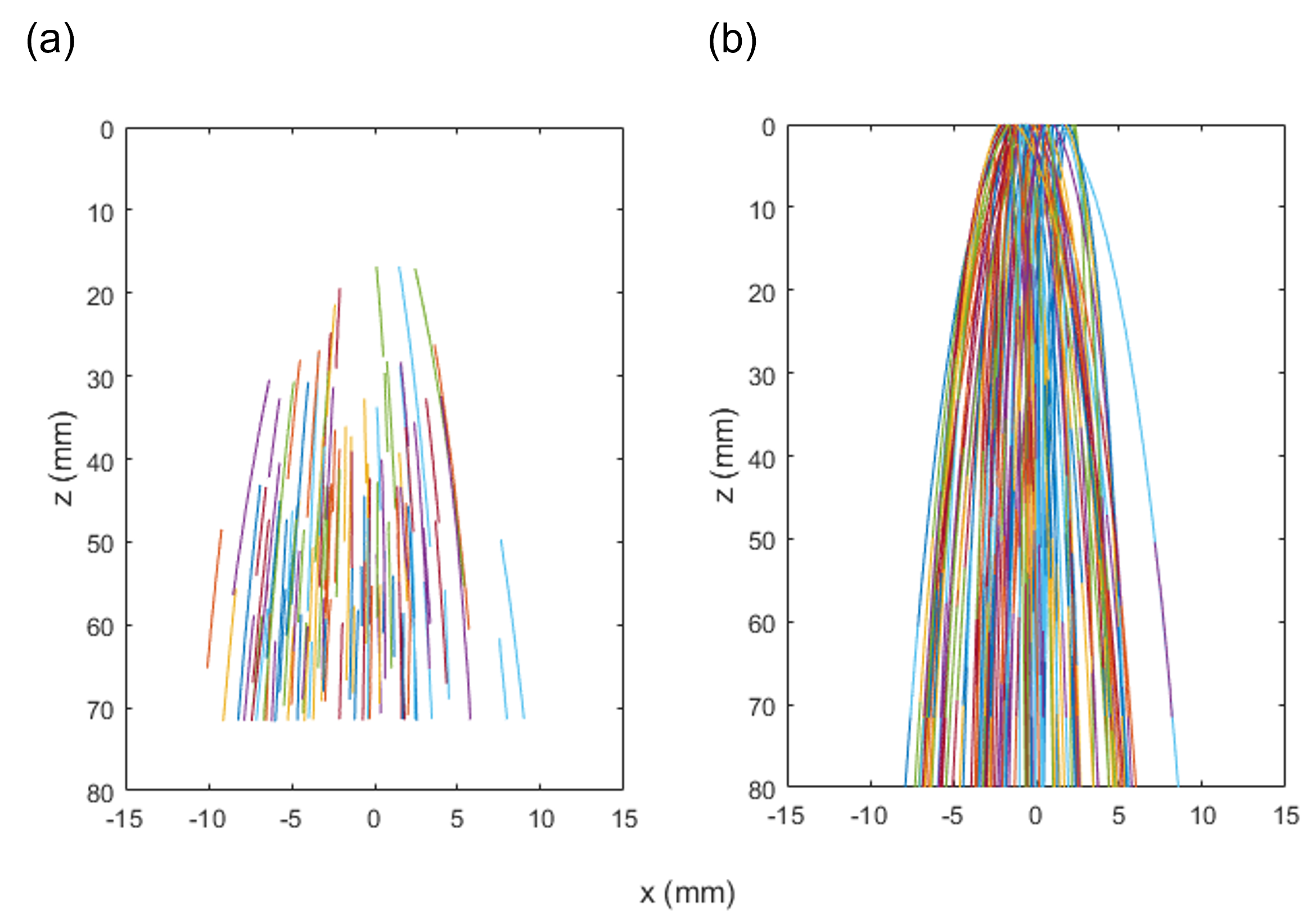}\\ 
\caption{Particle paths from (a) high speed video. (b) Extrapolated paths}
\vspace{0.5cm}
\label{fig:tracks}
\end{figure}

The filtered and extrapolated paths are shown in Fig.~\ref{fig:tracks}. Every 100 paths were plotted in Fig.~\ref{fig:tracks} (a) and every 20 paths were plotted in Fig.~\ref{fig:tracks} (b).

\begin{figure}[!ht]
\center
\vspace{0cm} 
\includegraphics[width=12cm]{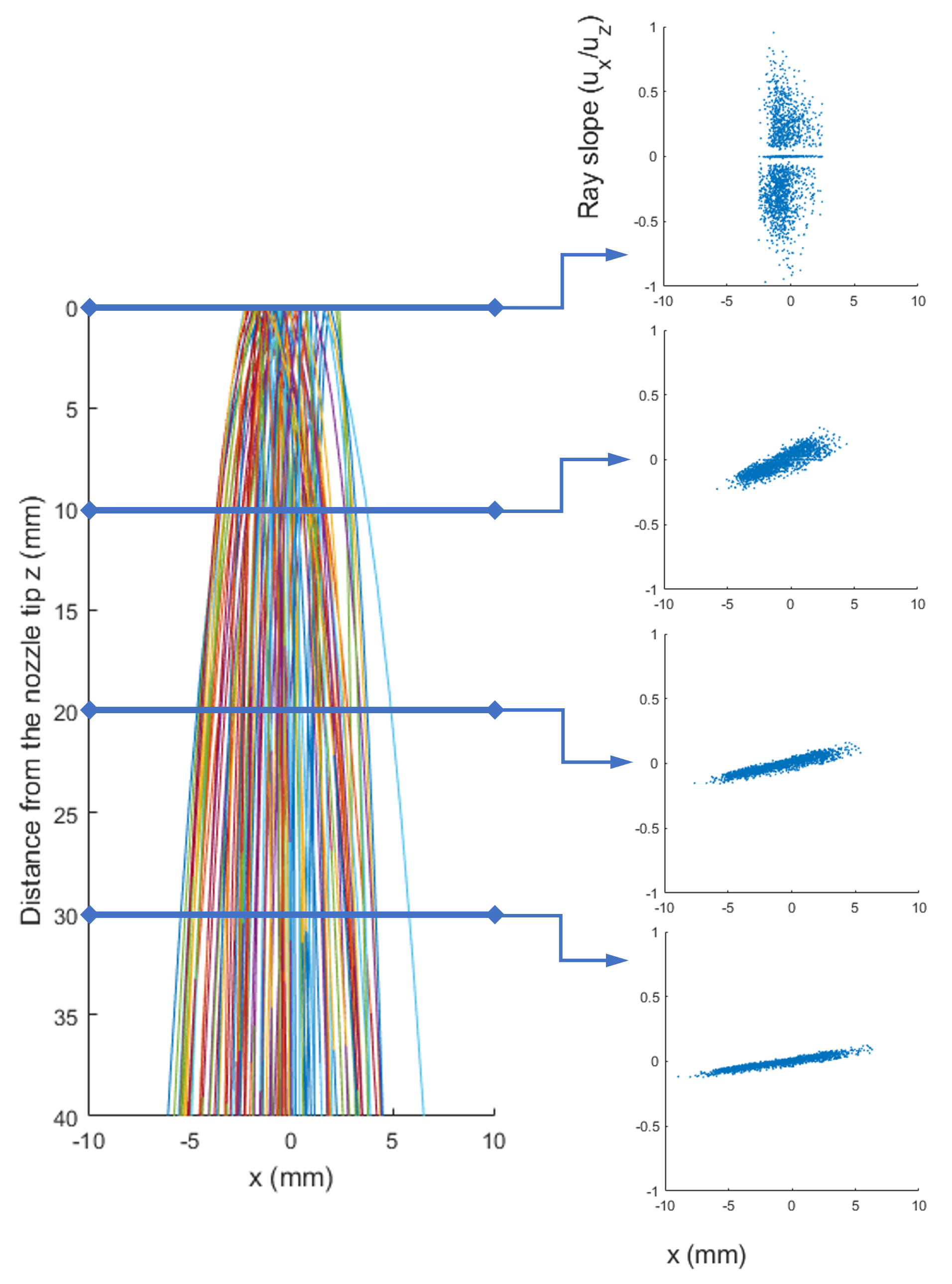}\\ 
\caption{Speed ratio of the particle paths as a function of $x$-location at different distances from the nozzle}
\vspace{0.5cm}
\label{fig:slopeScatterPlots}
\end{figure}

The speed ratio ($u_{0x}/u_{0z}$) of the particle paths as a function of $x$-location at different distances from the nozzle are given in Fig.~\ref{fig:slopeScatterPlots}. Note the decrease in spatial dependence as the paths approach the nozzle location. The increase of concentration data points at a speed ratio close to zero is due to fitting close to vertical paths with a straight line to prevent ill posed fitting are described in Section \ref{subsec:particleTracking}.

\begin{figure}[!ht]
\center
\vspace{0cm} 
\includegraphics[width=12cm]{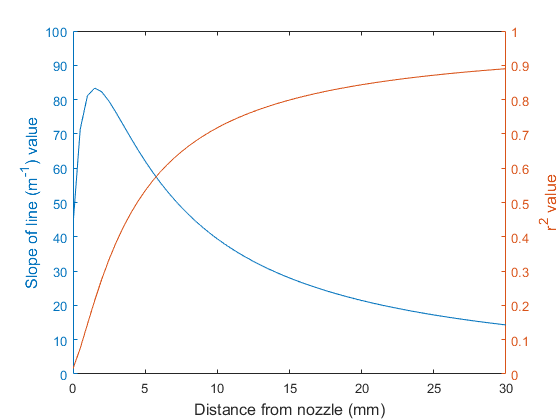}\\ 
\caption{Measured slope of speed ratio as a function of x and measured $r^2$ value for different nozzle distances}
\vspace{0.5cm}
\label{fig:measFitVarsOverZ}
\end{figure}

The slope of the fitted lines and their $r^2$ values along $z$ are shown in Fig.~\ref{fig:measFitVarsOverZ}. One can see a decrease in the $r^2$ value of the fit as the paths approach the nozzle location, showing a decrease in spatial dependence.

\begin{figure}[!ht]
\center
\vspace{0cm} 
\includegraphics[width=12cm]{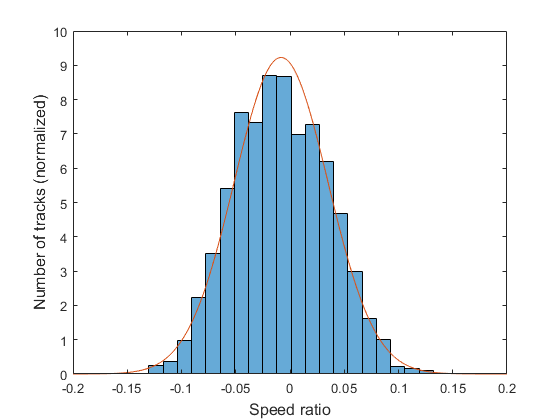}\\ 
\caption{Speed ratio of the particle paths downstream of applied force field}
\vspace{0.5cm}
\label{fig:speedRatioHistogram}
\end{figure}

The distribution of speed ratios at the $z$-value of interest downstream of the applied sound field is shown in Fig.~\ref{fig:speedRatioHistogram}. One can see that disregarding spatial dependence, the distribution is observed to be close to normal, with a corresponding $s_0$ (twice the standard deviation, see Eq.~\ref{eq:laserAngularDistPang}) of 0.0864. This value is not sensitive to reducing the number of trajectories: using trajectories within $x$ from -1.5 to 1.5 mm gives a value of 0.0860, and from -0.45 to 0.45 mm (corresponding to the nozzle diameter) gives a value of 0.0878. However, using these shorter ranges decreases the number of available trajectories for calculating the $r^2$ statistics.  

\subsection{Ray statistical characteristics assuming a Gaussian beam model}
\begin{figure}[!ht]
\center
\vspace{0cm} 
\includegraphics[width=12cm]{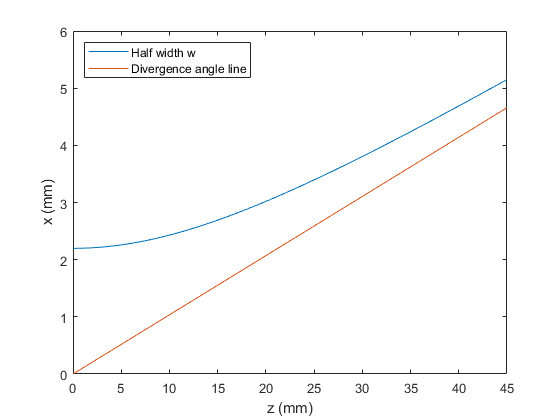}\\ 
\caption{Powder Gaussian beam model geometry}
\vspace{0.5cm}
\label{fig:powderGeometry}
\end{figure}

To account for gravitational acceleration, using the extrapolated path variables, one can calculate the average downward speed of the particles downstream of the applied force field and how long they take to reach that location from the nozzle location. This gives a distance of 44.2 mm. Then using the width and powder stream angle measured in \cite{martinezmarchese2022application}, Eqs.~\ref{eq:w0accurate} and \ref{eq:z0accurate} can be used to find $w_0$ and $z_R$. The $e^{-1}$ intensity values used are 3.59 mm for $w_1$ and 3.77\textdegree\ for $\theta_1$, note that these values should be multiplied by $\sqrt{2}$ to obtain the $e^{-2}$ intensity values. The values found for $w_0$ and $z_R$ were 2.2 mm and 21.2 mm respectively, with a ratio $s_0$ of 0.1035. This corresponds to the measured $s_0$ value from the path angles downstream of the sound field (0.0864), with an error of 19.8\%. The corresponding powder Gaussian beam geometry is shown in Fig.~\ref{fig:powderGeometry}.

\begin{figure}[!ht]
\center
\vspace{0cm} 
\includegraphics[width=12cm]{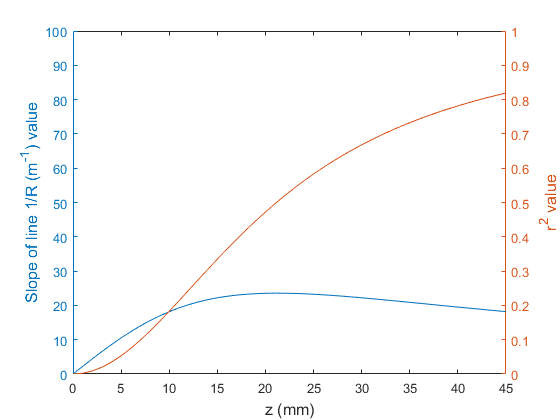}\\ 
\caption{Slope of $x$ vs. speed ratio and expected $r^2$ value of fitted line along $z$}
\vspace{0.5cm}
\label{fig:theoryFitVarsOverZ}
\end{figure}

The expected slope and $r^2$ value of the $x$-location vs. speed ratio along $z$ are shown in Fig.~\ref{fig:theoryFitVarsOverZ}. This plot has values quantitatively close to the ones in Fig.~\ref{fig:measFitVarsOverZ} at a $z$-value of 44.2 mm, while comparing it to the measured values at 27.5 mm. These values correspond to the same powder stream conditions downstream of the applied force field. The plot also has similar values at their corresponding nozzle position (0 mm), however the intermediate values are different especially for the speed ratio slopes; this is expected since the measured paths in Fig.~\ref{fig:measFitVarsOverZ} are subject to gravitational acceleration which `warps' the slope and $r^2$ values towards $x=\SI{0}{mm}$. One expects the speed ratio ($u_{0x}/u_{0z}$) to increase when $x$ approaches zero since the $u_{0z}$ values are expected to be lower at that location.

\subsection{Lagrangian particle simulation}
\label{sec:lagrangianParticleSimulation}
\begin{figure}[!ht]
\center
\vspace{0cm} 
\includegraphics[width=12cm]{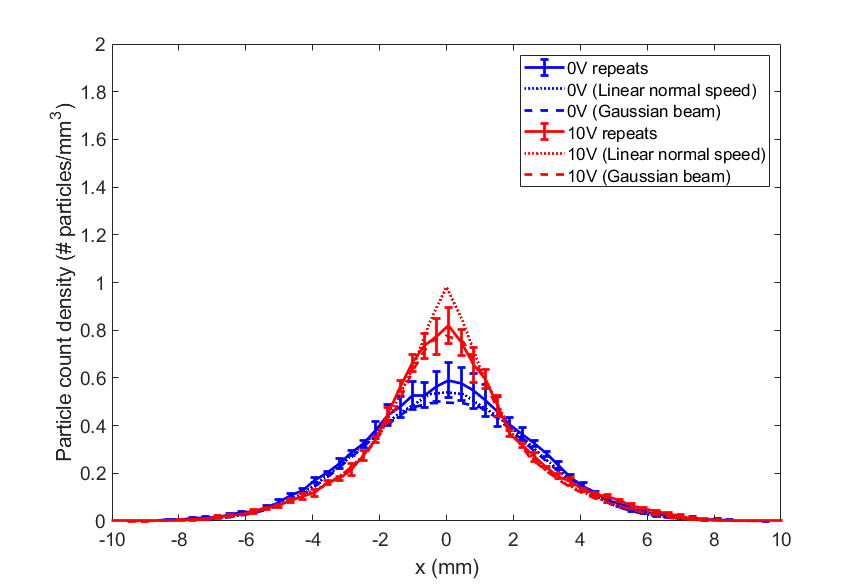}\\ 
\caption{Powder profile along $x$ downstream of the sound force field when using 0 and \SI{10}{V}}
\vspace{0.5cm}
\label{fig:concentrations10V}
\end{figure}

\begin{figure}[!ht]
\center
\vspace{0cm} 
\includegraphics[width=12cm]{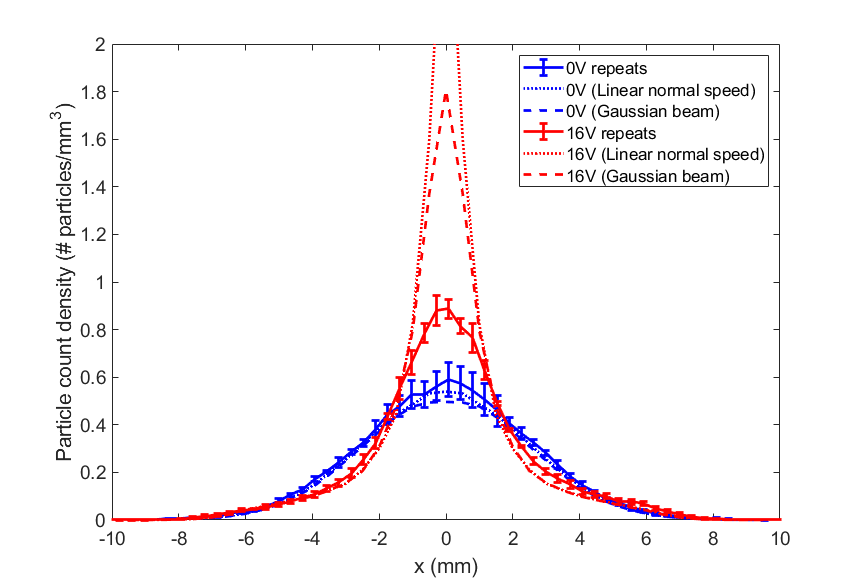}\\ 
\caption{Powder profile along $x$ downstream of the sound force field when using 0 and \SI{16}{V}}
\vspace{0.5cm}
\label{fig:concentrations16V}
\end{figure}

The calculated powder profile along the $x$ axis downstream of the sound force field when using \SI{10}{V} with the one at \SI{0}{V} is shown in Fig.~\ref{fig:concentrations10V}. For the force field corresponding to \SI{0}{V}, since the paths extend to the $z$-location with no force being applied, there is a close match between the measured and simulated data as expected. There is also a good agreement with the simulated results at \SI{10}{V}. The profile at \SI{16}{V} with the one at \SI{0}{V} is shown in Fig.~\ref{fig:concentrations16V}. The peak PCD for \SI{16}{V} of \SI{1.8}{particles/mm^3} is still higher than the measured value of \SI{0.9}{particles/mm^3}, (a 100\% difference) but it is still much lower than the result of \SI{2.9}{particles/mm^3} (a 222\% difference) when assuming a deterministic linearly increasing outward velocity component as assumed in \cite{martinezmarchese2022application}. Other factors causing this discrepancy could be due to particle collisions and/or air drafts present when recording the high-speed data, which had to be minimized with an enclosure \cite{martinezmarchese2022application}.

\begin{figure}[!ht]
\center
\vspace{0cm} 
\includegraphics[width=12cm]{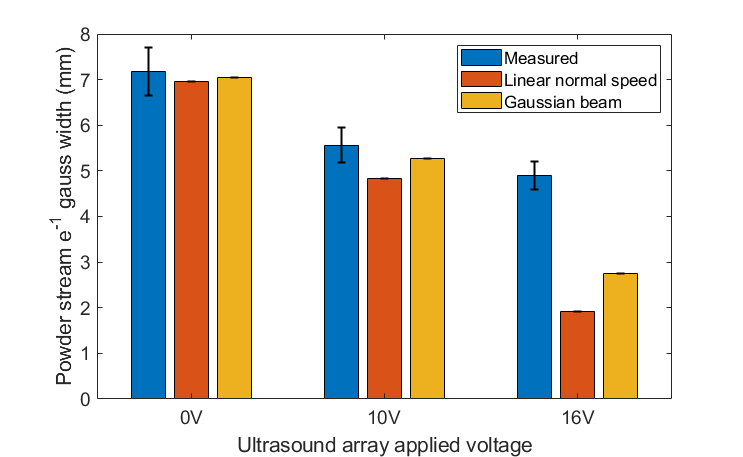}\\ 
\caption{Comparison of measured powder stream widths and simulated powder stream widths}
\vspace{0.5cm}
\label{fig:concentrationWidths}
\end{figure}

A comparison of the $e^{-1}$ powder widths are shown in Fig.~\ref{fig:concentrationWidths}. As expected, the closer PCD profiles seen in Figs.~\ref{fig:concentrations10V} and \ref{fig:concentrations16V}, produce a closer match between the measured and calculated powder widths for all voltages when using the Gaussian beam model, compared to the linearly increasing outward velocity component model in \cite{martinezmarchese2022application}.

\begin{figure}[!ht]
\center
\vspace{0cm} 
\includegraphics[width=12cm]{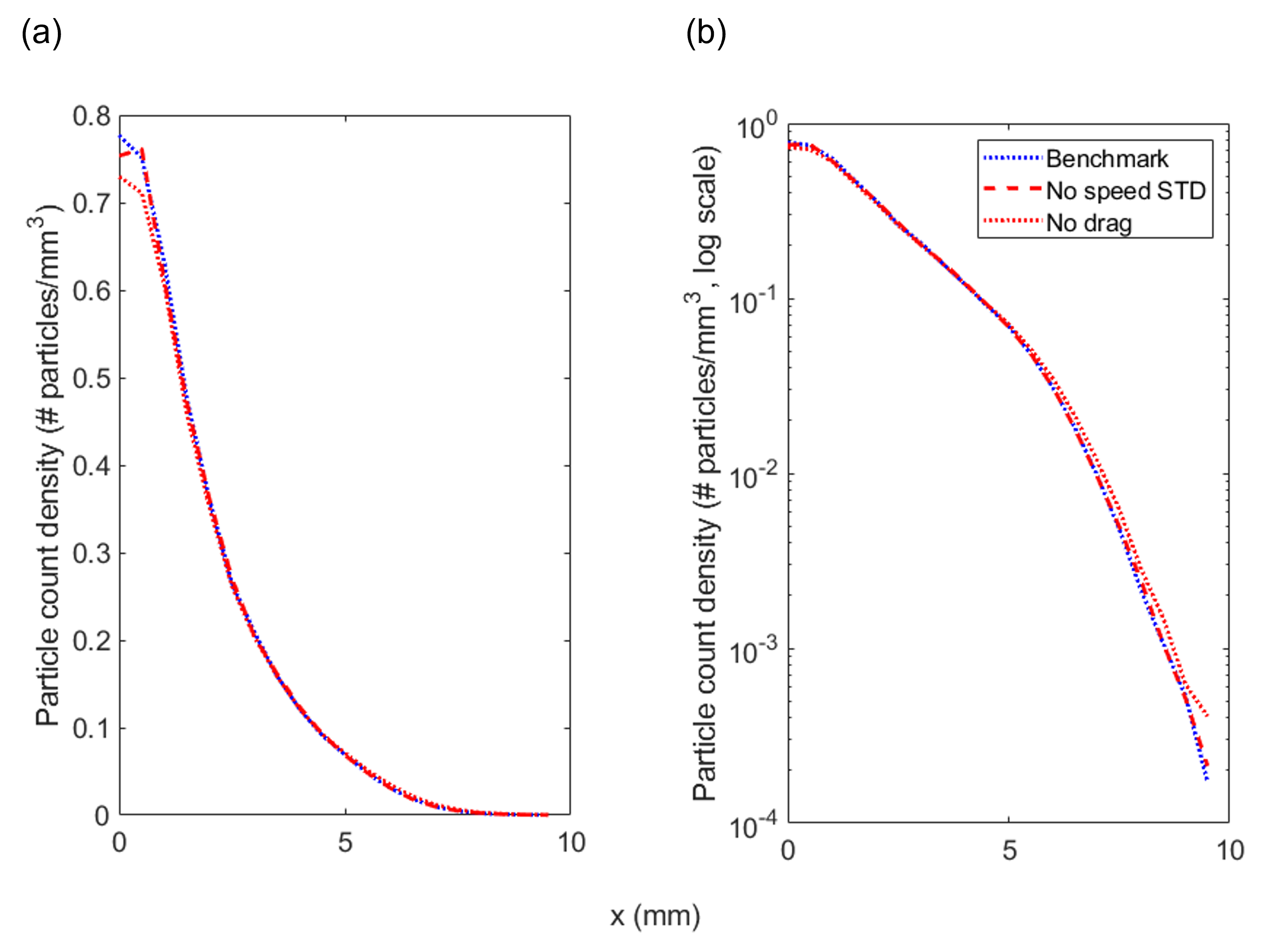}\\ 
\caption{Powder profile along $x$ downstream of the sound field considering STD and air drag, only considering drag or considering only speed standard deviation. Using a $y$-axis linear scale (a) and a logarithmic scale (b)}
\vspace{0.5cm}
\label{fig:simComparison}
\end{figure}

Using the Lagrangian simulation, one can check the effect of each phenomenon  modeled on the final particle concentration, as shown in Fig.~\ref{fig:simComparison}. There is no clear difference seen in Fig.~\ref{fig:simComparison} (a) between the full Lagrangian simulation and the simulation without a standard deviation of the particle speed or without drag, except for the expected 4\% variation at the concentration peak due to the number of particle paths used in the MC model. However, there is a slight decrease in the peak concentration above 4\% for the case of no drag. This is shown more clearly using a log scale for the $y$-axis, as shown in Fig.~\ref{fig:simComparison} (b), where there is an increase in the concentration in the tails of the PCD. This is explained by the fact that with no drag force, the particles are expected to focus less due to less drag force applied to the particles in the same direction as the focusing force field. For the speed standard deviation measured, the particle size density, particle speeds, and the 20 mm sound interaction length being considered in this setup, the effect of the standard deviation and drag is not significant.

The speed variation might not be a significant factor in the shape of the PCD for the following reason; although the particle's deviation in a focusing force field approximately depend on their downward speed squared \cite{martinezmarchese2022application}, the change in deviation will decrease towards the center of the powder stream, where the particles are not affected by the field. If a particle that is moving on the powder stream center-line, has a deviation in its speed, and it is not affected by the force field, it will reach the same location regardless of the speed variations.

\subsection{Powder focusing modeled as an optical system}
\begin{figure}[!ht]
\center
\vspace{0cm} 
\includegraphics[width=13cm]{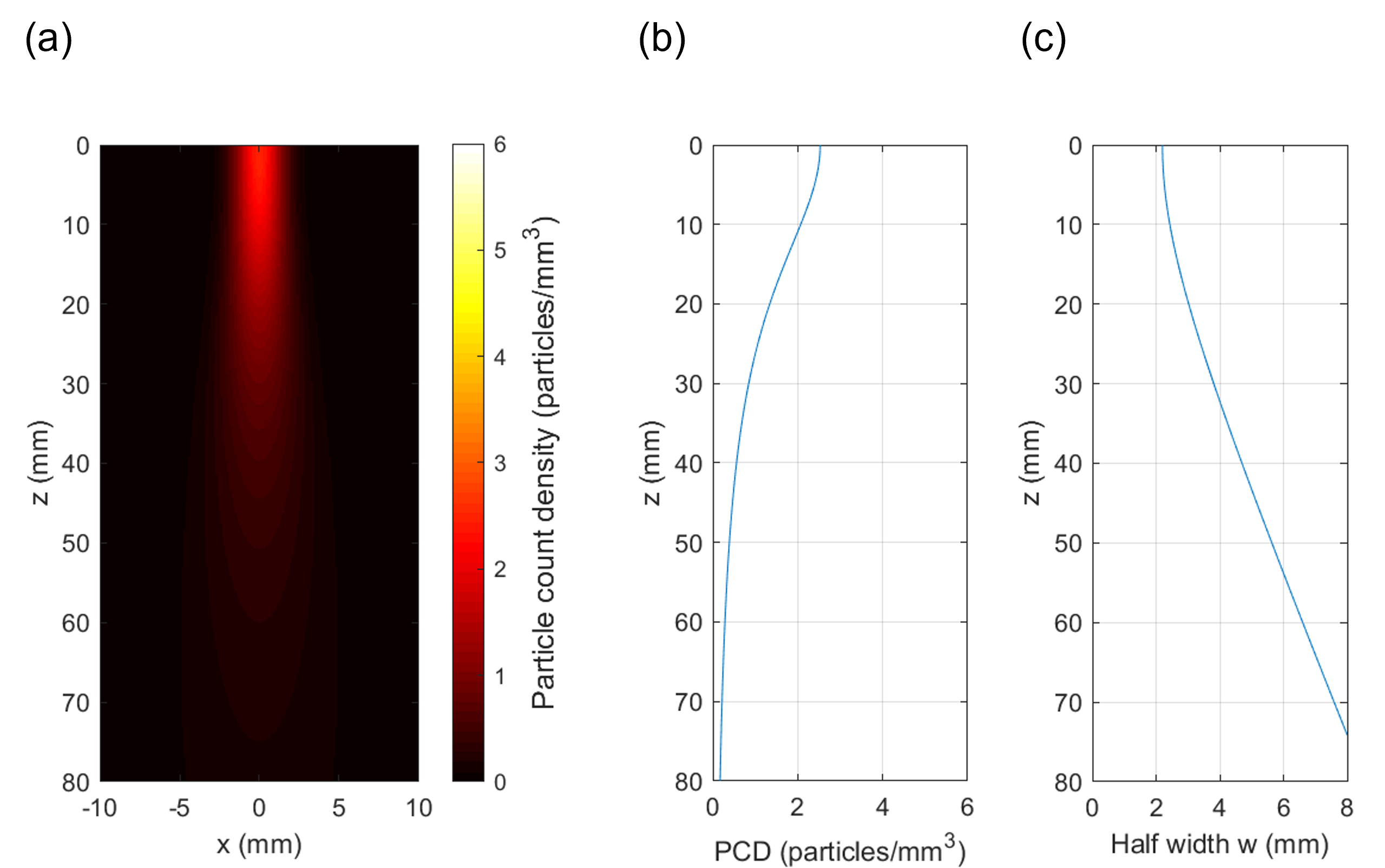}\\ 
\caption{(a) Powder profile, (b) concentration at center-line along $z$ and (c) half width at center-line along $z$, all at \SI{0}{V}}
\vspace{0.5cm}
\label{fig:rayTracing0V}
\end{figure}

\begin{figure}[!ht]
\center
\vspace{0cm} 
\includegraphics[width=9cm]{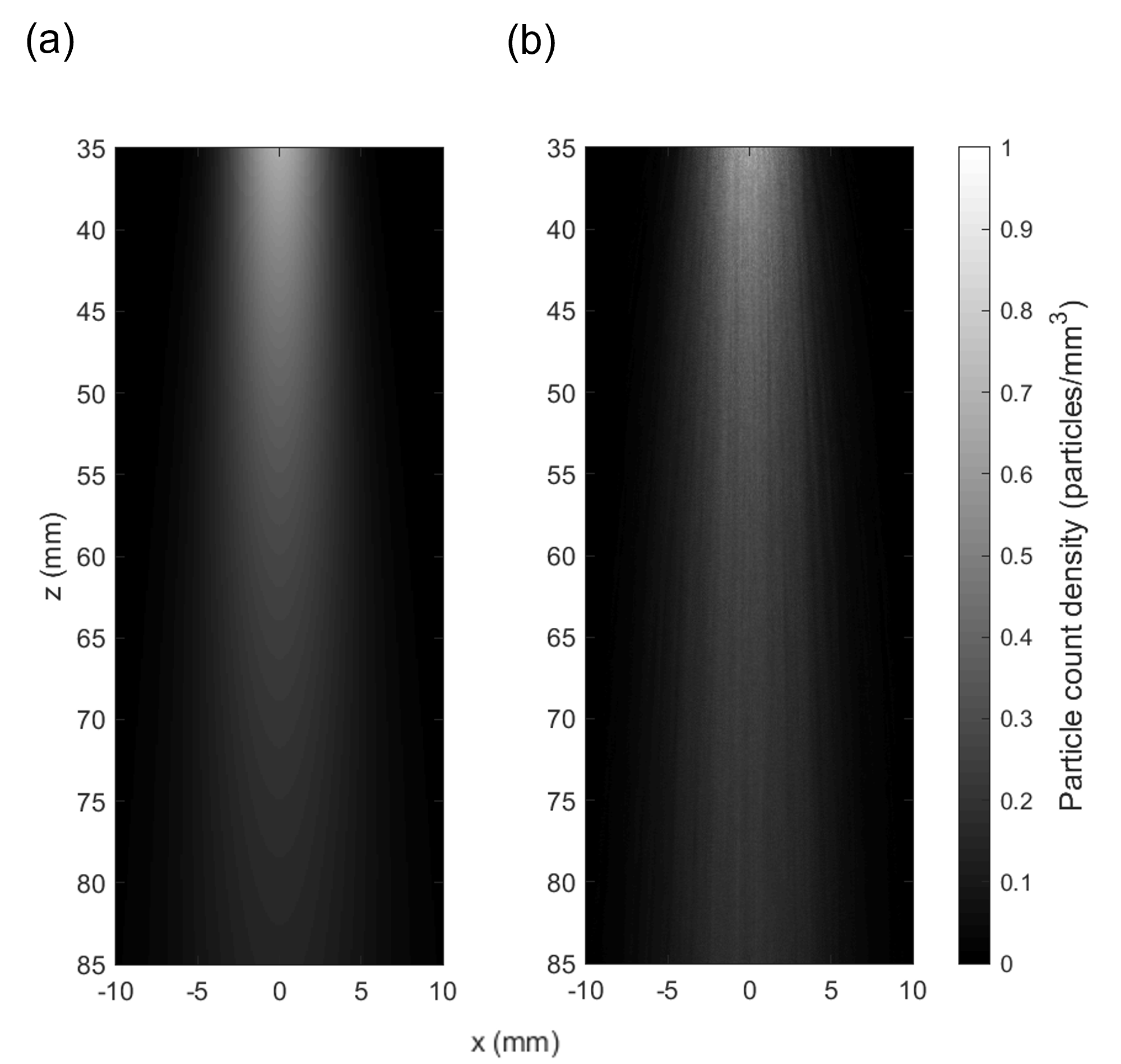}\\ 
\caption{Calculated and measured powder profiles at \SI{0}{V}}
\vspace{0.5cm}
\label{fig:rayTracingPred0V}
\end{figure}

The calculated powder profile and width along $z$ when an external sound force field using \SI{0}{V} is applied is shown in Fig.~\ref{fig:rayTracing0V}. As seen in Fig.~\ref{fig:rayTracing0V}~(c), the width increases linearly along the $z$-axis at large values of z, as suggested in the DED-PF literature \cite{huang2016comprehensive}. The calculated powder profile with the measured powder profile is shown in Fig.~\ref{fig:rayTracingPred0V}. The recorded powder profile is in close agreement with the simulated profile up to about $z = \SI{35}{mm}$ but there is less agreement for higher values of $z$. This might be due to gravitational acceleration of the particles. 

\begin{figure}[!ht]
\center
\vspace{0cm} 
\includegraphics[width=13cm]{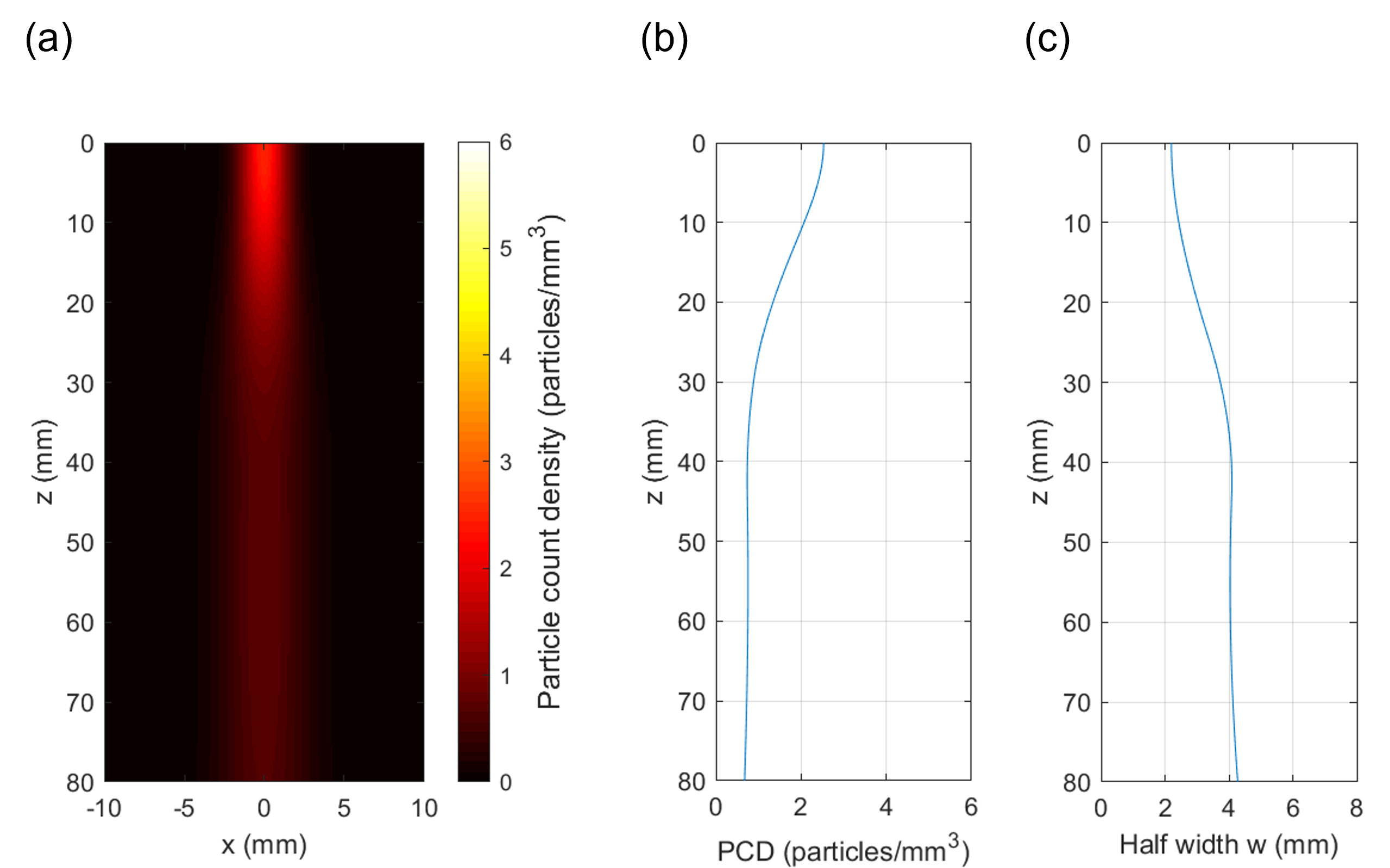}\\ 
\caption{Powder profile (a), concentration at center-line along $z$ (b) and half width at center-line along $z$ (c) at \SI{10}{V}}
\vspace{0.5cm}
\label{fig:rayTracing10V}
\end{figure}

\begin{figure}[!ht]
\center
\vspace{0cm} 
\includegraphics[width=9cm]{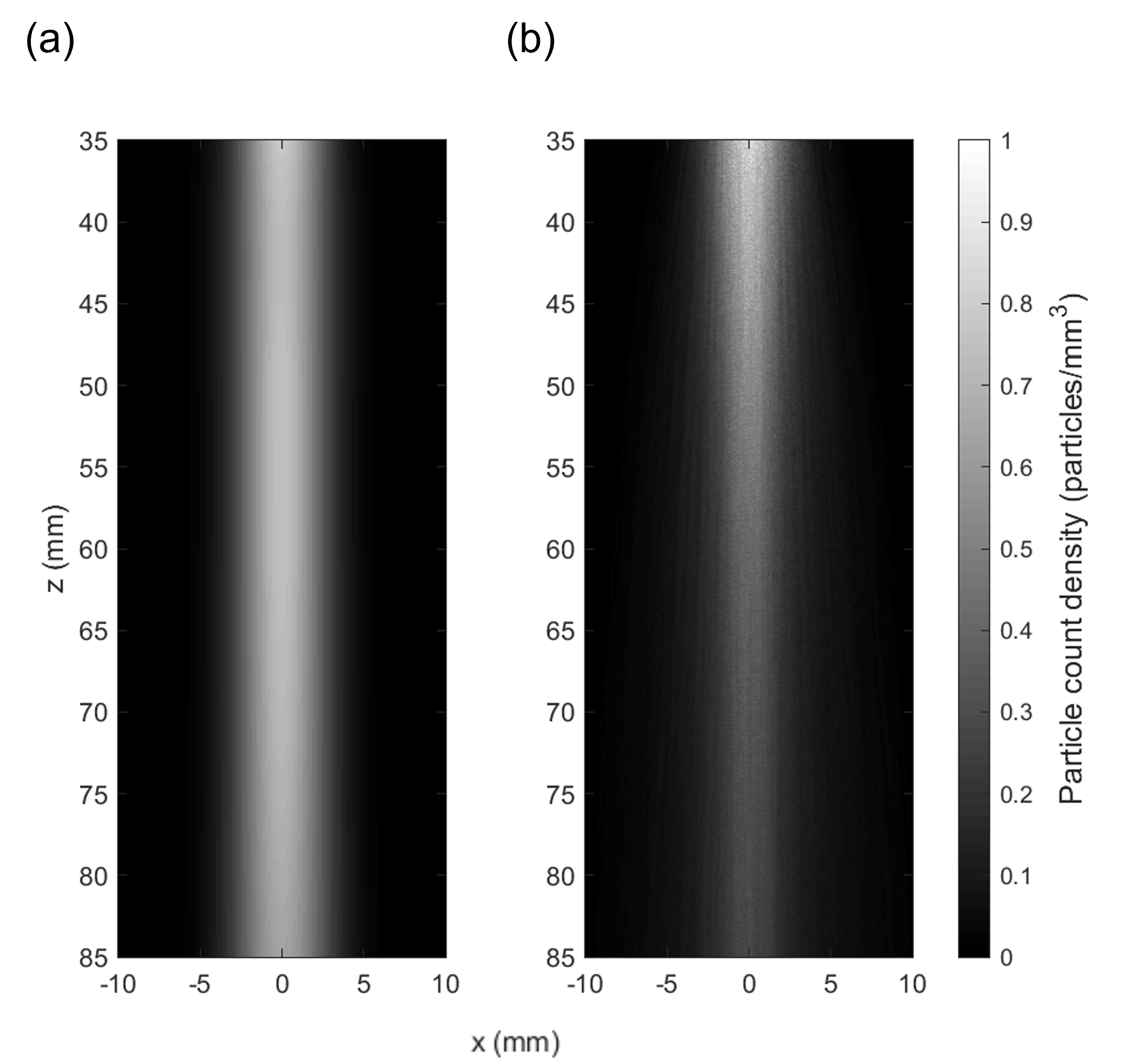}\\ 
\caption{Calculated and measured powder profiles at \SI{10}{V}}
\vspace{0.5cm}
\label{fig:rayTracingPred10V}
\end{figure}

The calculated powder profile and width along $z$ at \SI{10}{V} is shown in Fig.~\ref{fig:rayTracing10V}. As seen in this model with this force field, one can predict a `shallow' focus, where the width and particle concentration change slowly along $z$, corresponding to the particle's angles with respect to the powder center-line on average being close to zero. This is seen in Figs.~\ref{fig:rayTracing10V} (b) and (c). The calculated powder profile is compared with the measured powder profile as shown in Fig.~\ref{fig:rayTracingPred10V}. For this force field strength corresponding to \SI{10}{V} there is close agreement between the simulated and measured data. Note that since particles are moving on average close to parallel to the powder stream center-line, the effect of gravitational acceleration should be less noticeable.

\begin{figure}[!ht]
\center
\vspace{0cm} 
\includegraphics[width=13cm]{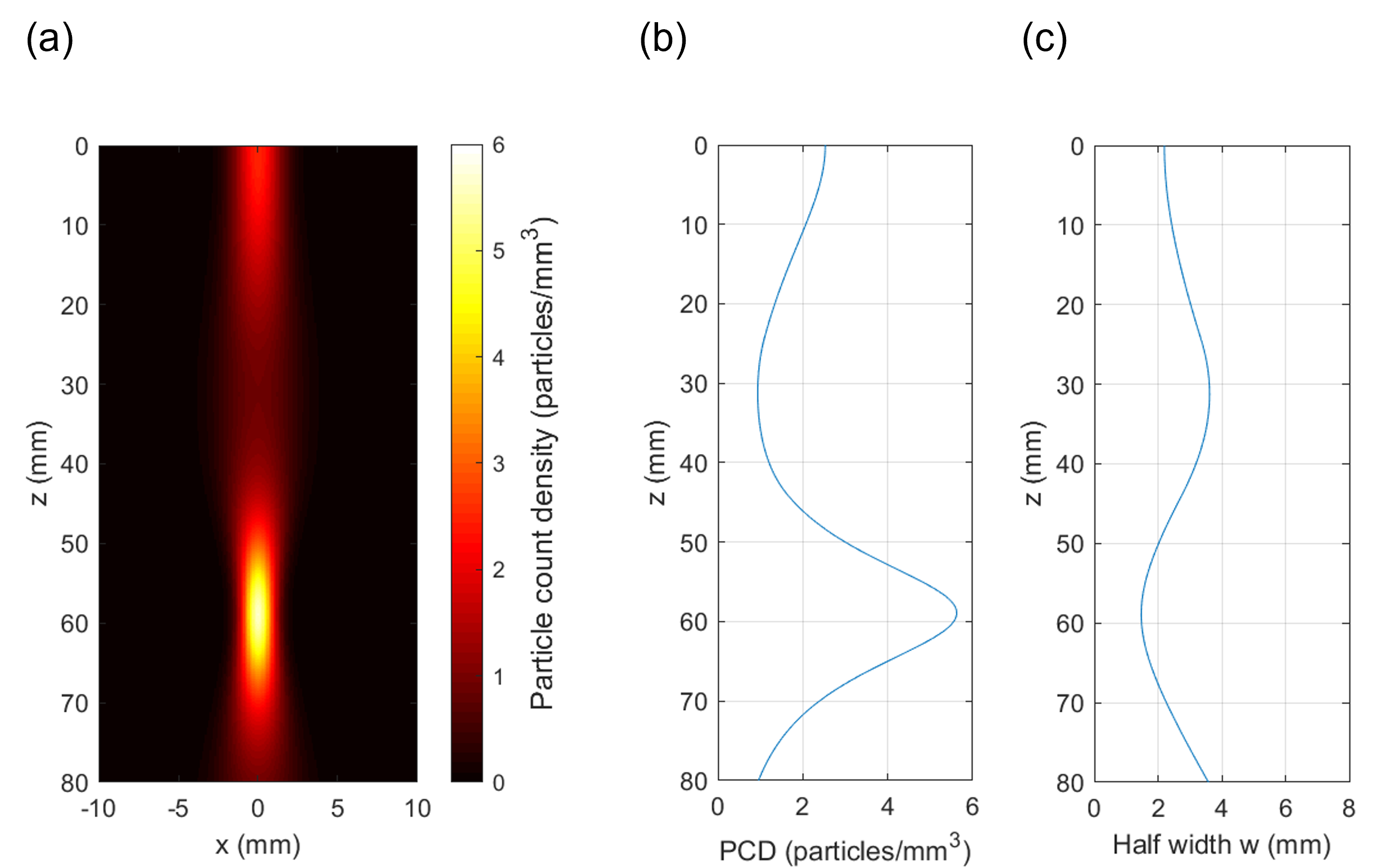}\\ 
\caption{Powder profile (a), concentration at center-line along $z$ (b) and half width at center-line along $z$ (c) at \SI{16}{V}}
\vspace{0.5cm}
\label{fig:rayTracing16V}
\end{figure}

\begin{figure}[!ht]
\center
\vspace{0cm} 
\includegraphics[width=9cm]{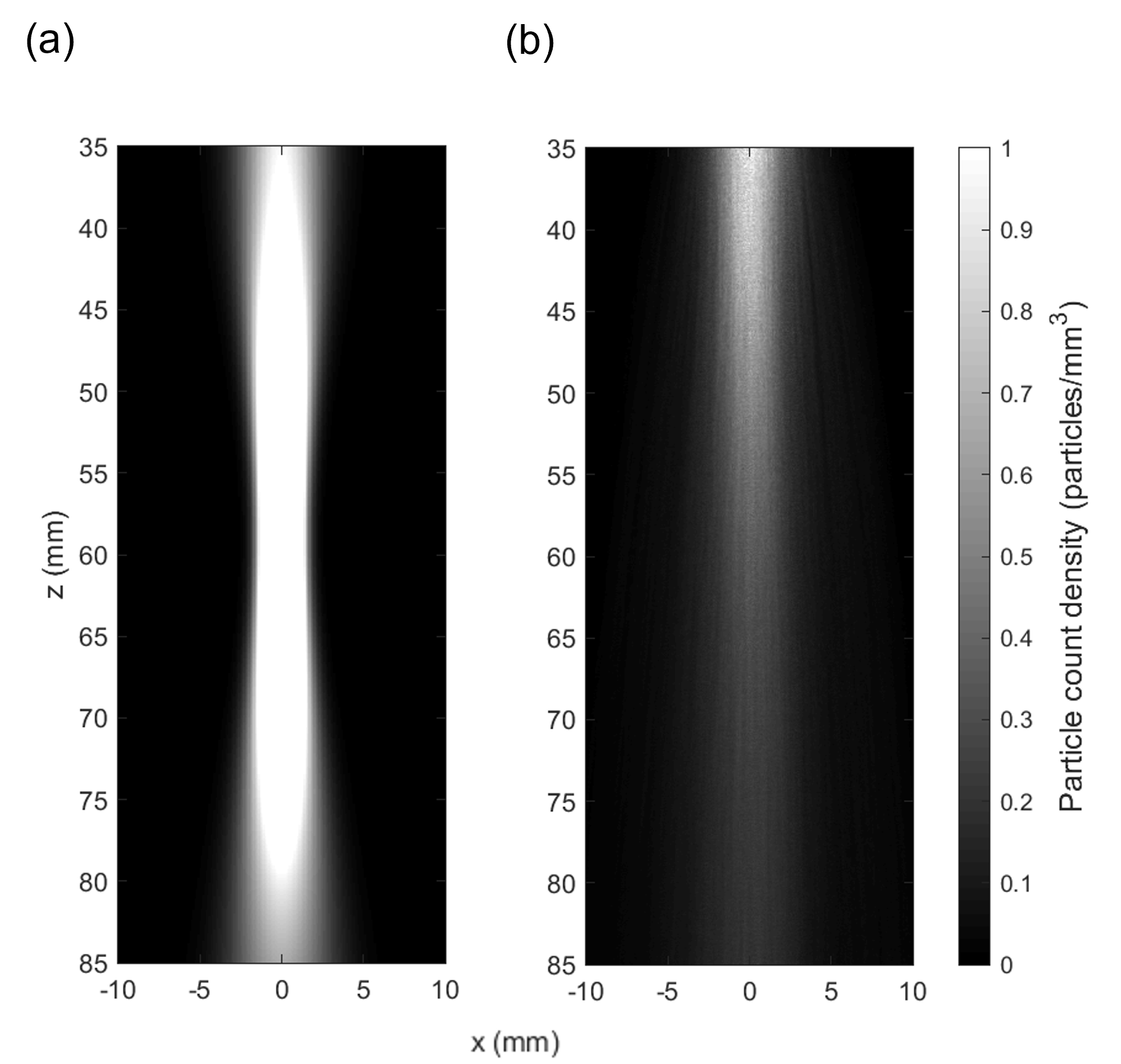}\\ 
\caption{Calculated and measured powder profiles at {\SI{16}{V}}}
\vspace{0.5cm}
\label{fig:rayTracingPred16V}
\end{figure}

The calculated powder profile along $z$ at \SI{16}{V} is shown in Fig.~\ref{fig:rayTracing16V}. The calculated powder profile is compared with the measured powder profile in Fig.~\ref{fig:rayTracingPred16V}. The model predicts a sharp increase in concentration to \SI{5.6}{particles/mm^3} at $z =\SI{59}{mm}$. Comparing this result with the experimental data in Fig.~\ref{fig:rayTracingPred16V}~(b), the concentration peaks at about \SI{1}{particle/mm^3} and decrease downstream. The main cause of this discrepancy is the fact that the experiment carried out in \cite{martinezmarchese2022application} had an initial $e^{-1}$ powder stream width of $\SI{7.2}{mm} - $2(20.0)tan(3.77\textdegree) mm = $\SI{2.6}{mm}$,\linebreak compared to a force offset width of about \SI{2}{mm} \cite{martinezmarchese2022application}, corresponding to the distance from the powder stream center-line where the force field can be assumed to be linear. This means that the GRIN lens model is overestimating the amount of focusing for particles farther away than 1 mm from the center-line, producing a higher concentration at the focus point. Other factors causing this discrepancy could be due to the small-angle tangent approximation used for the GRIN lens analog model (Eq.~\ref{eq:ABCDgrin}), particle collisions and/or air drafts present when recording the high-speed data \cite{martinezmarchese2022application}.

A summary of the obtained parameters for the three simulations is shown in Table \ref{table:opticAnalogResults}. The peak concentration was the maximum concentration after $z = \SI{50}{mm}$ and the powder width location is defined as the distance from downstream of the force field to the powder `beam' waist. Note that the model indicates that using 16 V one may achieve a waist width smaller than the initial width $2w_0$ of 4.4 mm. The peak concentrations downstream of the force field ($z =\SI{44.2}{mm}$) shown on Table \ref{table:opticAnalogResults} closely agree with the Lagrangian model, with a percent error with respect to the Lagrangian model that may be due to the expected error due to the random sampling of the particle paths or because a speed standard deviation and air drag is considered in the Lagrangian model.

\begin{table}[ht]
\footnotesize
\centering
\vspace{0.4cm}
\caption{Predicted powder stream parameters from optical analog computations and percent error peak particle concentration comparison with Lagrangian model}
\begin{tabular}
{>{\raggedright\arraybackslash}p{1.5cm}>{\raggedright\arraybackslash}p{1.5cm}>{\raggedright\arraybackslash}p{1.5cm}>{\raggedright\arraybackslash}p{1.5cm}>{\raggedright\arraybackslash}p{2.5cm}>{\raggedright\arraybackslash}p{2.5cm}>{\raggedright\arraybackslash}p{2.0cm}}
\hline\hline
\\
\textbf{Applied voltage (V)} & \textbf{Force GRIN lens pitch} & \textbf{Powder waist location (mm)} & \textbf{Powder waist width (mm)} & \textbf{Peak concentration (\SI{}{particles/mm^3})} & \textbf{Concentration downstream of force field (\SI{}{particles/mm^3})} & \textbf{\% error w.r.t Lagrangian simulation} \\
& \\
\hline
& \\
0 & 0.0 & NA & NA & NA & 0.47 & -4.4 \newline \\

10 & 0.12 & 11.0 & 8.1 & 0.75 & 0.73 & -5.9 \newline \\

16 & 0.19 & 14.7 & 3.0 & 5.6 & 1.65 & -8.9 \\

& \\
\hline
\end{tabular}
\vspace{0.4cm}
\label{table:opticAnalogResults}
\end{table}

As an example of the powder waist width that may be achieved using a particular force field, using the same $w_0$ and $z_R$ values used in the results, Eq.~\ref{eq:lfs} to find $l_{fs}$ and then using this value to find $\eta$ such that the numerator of Eq.~\ref{eq:zsAnalytic} equals to zero, one obtains the values shown in Fig.~\ref{fig:yyBarDiags} (c). The $\eta$ was found using the function \verb|fzero()| is MATLAB, with an initial guess of 70. The assumed $k$ value was 1 and the force field radius was assumed to be 3 mm. This produces a waist width of 1.6 mm at the force field boundary, which is 2.75 times smaller than $w_0$, equal to 4.4 mm. This corresponds to a GRIN pitch analog of 0.29, a peak particle concentration of \SI{19.1}{particles/mm^3}. The $\eta$ value corresponds to an ultrasound array applied voltage of 25 V, assuming that the particle applied force $F_a$ for the definition of $\eta$ is proportional to the applied voltage squared \cite{martinezmarchese2022application}, and the $\eta$ value is 59.4 m\textsuperscript{-1} at 16 V.

\section{Conclusions}
A model that more accurately predicts the behavior of a powder stream, with similar characteristics to the ones in DED-PF, when subject to a focusing force field was described. The model uses a Gaussian distribution for the initial particle positions, and a Gaussian distribution for the normal component of the initial particle velocities. 
The following conclusions can be drawn from this work and the Gaussian ray representation applied to powder stream modeling:
\begin{itemize}
    \item The model closely matches the initial conditions seen in the extrapolated particle tracks that are close to the nozzle, with normally distributed normal velocity components, and close to zero spatial dependence. The sideways speed standard deviation calculated using particle trajectories and assuming the novel Gaussian beam ray model, only considering the final PCD from all the particles, match within 20\%
    \item Both the slope of the angle of the trajectories with respect to the $z$ axis as a function $x$ and the $r^2$ values measured from the trajectories and derived from the model are similar towards larger $z$ values (Figs.~\ref{fig:measFitVarsOverZ} and \ref{fig:theoryFitVarsOverZ}) when the effect of gravitational  acceleration is less pronounced
    \item The model follows mass conservation
    \item At values far from the nozzle the IRW in this model approaches a linear function with respect to $z$, as seen in the experimental data
    \item The model also predicts a parabolic function of the form $w = w_0 + a z^2$, where $a$ is a fitted constant, for the region close to the powder nozzle as observed in \cite{toyserkani2004laser}. This can be seen by computing the first two terms of the series expansion of Eq.~\ref{eq:w} at $z=0$, yielding $w \approx w_0 [1 + z^2/(2 z_R^2)]$
    \item The model can also be used to predict a minimum achievable powder spot width when an external symmetric force field is applied to the powder stream. It can be used to more quickly compute the PCD, using an optical system analog and can be used to predict what factors in the DED-PF powder feed affect the minimum powder spot width
    \item When the assumption of of a normally distributed sideways particle speed is implemented as a Lagrangian model, it predicts with more accuracy the PCD downstream of the field and the observed $e^{-1}$ intensity widths (Figs.~\ref{fig:concentrations10V} to \ref{fig:concentrationWidths})
    \item The above Lagrangian model with different force field magnitudes was then compared with the novel Gaussian beam ray model; the percent error of the peak particle concentration downstream of the force field between the Lagrangian model and the Gaussian beam ray model was below 6 \% at 0 and 10 V and below 9 \% at 16 V (Table~\ref{table:opticAnalogResults})
    \item The calculation of the PCD using the new Gaussian beam model takes 0.2 seconds in MATLAB, compared to a minimum of 34 minutes (16 V case) for the converged Lagrangian simulation written in C++
    \item Both Lagrangian and Gaussian beam ray simulations still deviate from the measured concentrations at higher forces (corresponding to \SI{16}{V}); this might be due to inaccuracies in the measured values used to fit the model, particle collisions and/or air drafts present when recording the high speed data, which had to be minimized with an enclosure \cite{martinezmarchese2022application}. In the case of the laser propagation model, the inaccuracy of the model for large applied force fields (when using \SI{16}{V}) might be related to the large width of the powder stream upstream of the sound with respect to the linear range of the sound assumed in the analytic derivation of the particle paths done in \cite{martinezmarchese2022application}. This discrepancy could be reduced when simulating an initially narrower powder stream or force field with a different radial profile. More experimental data and a Lagrangian model with particle collisions could be used to investigate this hypothesis
    \item The Lagrangian simulation was used to determine that Gaussian speed variations and air drag (only due to the force field moving the particles) do not significantly affect the PCD for the speed STD and divergence angle measured
    \item Better simulation results are expected when the initial powder stream width is smaller than the force field offset width \cite{martinezmarchese2022application}. Better predictions should also be possible when modeling powder streams with higher speed particles, normally used in DED-PF equipment, where the effect of particle gravitational acceleration is less significant. In this case one could also obtain the $w_0$ and $z_0$ parameters using two stream widths at two different $z$-values instead of using Eqs.~\ref{eq:w0accurate} and \ref{eq:z0accurate}
\end{itemize}

Modeling the powder stream as a Gaussian beam, besides allowing the simulation of focusing from a radially symmetric force field, could also be used to model powder stream spot shape changes due to an non-symmetric force field, using the tensor ABCD law \cite{lin1990transformation}, and powder stream small angle deflection with a force field, using a decentered (with respect to a GRIN lens corresponding to an off-center vortex sound force field for example) Gaussian beam formalism \cite{cai2002decentered}. It may also be possible to use a coordinate transformation along the $z$ axis in order to take into account gravitational acceleration.

The powder stream model described in this article could be used for more accurate DED-PF melt pool simulations that include a powder stream \cite{bayat2021role} and modeling a laser beam with the Gaussian beam ray representation could be used in simulations of laser interaction with metal powder in the laser powder bed fusion process \cite{kovalev2018ray, devesse2015modeling}.

\section{CRediT authorship contribution Statement}
A. Martinez-Marchese: Conceptualization, Methodology, Software, Validation, Formal analysis, Investigation, Writing – original draft, Writing – review \& editing, Visualization. M. Klumpp: Formal analysis, Writing – review \& editing. E. Toyserkani: Conceptualization, Writing – review \& editing, Supervision, Project administration, Funding acquisition.

\section{Declaration of Competing Interest}
The authors declare that they have no known competing financial interests or personal relationships that could have appeared to influence the work reported in this paper.

\section{Acknowledgments}
The authors would like to acknowledge the financial support of the Natural Sciences and Engineering Research Council of Canada (NSERC) Network on Holistic Innovation in Additive Manufacturing (HI-AM)

\bibliographystyle{ieeetr}

\begingroup
\raggedright
\bibliography{main}
\endgroup

\newpage
\appendix
\section{Calculation of parameters for the expected value of $r^2$ (Eq.~\ref{eq:R2})}
\setcounter{equation}{0}
\renewcommand\theequation{A.\arabic{equation}}

Here we want to calculate $\mathbb{E}(SS_{res})/\mathbb{E}(SS_T)$ from Eqs.~\ref{eq:ESSres}-\ref{eq:ESST}. Note that:

\begin{align*}
    \frac{\mathbb{E}(EE_{res})}{\mathbb{E}(EE_{T})}=\frac{N}{N-1}\frac{\mathbb{E}(y_n^2)-\mathbb{E}(\mathbb{E}_{\theta_x}^2(y_n))}{\mathbb{E}(y_n^2)-\mathbb{E}^2(y_n)}\overset{N\rightarrow\infty}{\rightarrow}\frac{\mathbb{E}(Y^2)-\mathbb{E}(\mathbb{E}_{\theta_x}^2(Y))}{\mathbb{E}(Y^2)-\mathbb{E}^2(Y)}
\end{align*}
\hfill \break
For this we used:

\begin{align}
    \mathbb{E}\left((Y-\mathbb{E}_{\theta_x}(Y))^2\right) &= \mathbb{E}\left(Y^2\right)-2\cdot\mathbb{E}(Y\cdot\mathbb{E}_{\theta_x}(Y))+\mathbb{E}(\mathbb{E}_{\theta_x}^2(Y))=\mathbb{E}_x(\mathbb{E}_{\theta_x}(Y^2))-\mathbb{E}_x(\mathbb{E}_{\theta_x}^2(Y)) \label{eq:nominator}\\
    \mathbb{E}\left((Y-\mathbb{E}(Y))^2\right) &= \mathbb{E}\left(Y^2\right)-2\cdot\mathbb{E}(Y\cdot\mathbb{E}(Y))+\mathbb{E}(\mathbb{E}^2(Y))=\mathbb{E}_x(\mathbb{E}_{\theta_x}(Y^2))-\mathbb{E}_x^2(\mathbb{E}_{\theta_x}(Y)) \label{eq:denominator}
\end{align}
\hfill \break
Therefore $\mathbb{E}_{\theta_x}(Y)$ and $\mathbb{E}_{\theta_x}(Y^2)$ need to be calculated first:

\begin{align*}
    \mathbb{E}_{\theta_x}(Y)&\overset{Eq.~\ref{eq:Ethetax}}{=}\int_{-\frac{\pi}{2}}^{\frac{\pi}{2}}\theta_x\cdot f_{\theta_x}(\theta_x|x)\,d\theta_x\\
    &\overset{Eq.~\ref{eq:condDensityAngleXFreeProp}}{=}\int_{\mathbb{R}}\frac{\arctan\left(\frac{k_x}{k}\right)}{\sqrt{2\pi}k_1(z)}\exp\left(-\frac{\left(k_x-k\frac{x}{R(z)}\right)^2}{2k_1^2(z)}\right)\,dk_x\\
    &=\sum_{n=0}^\infty\frac{(-1)^n}{\sqrt{2\pi}k_1(z)(2n+1)k^{2n+1}}\int_\mathbb{R}k_x^{2n+1}\exp\left(-\frac{\left(k_x-k\frac{x}{R(z)}\right)^2}{2k_1^2(z)}\right)\,dk_x
\end{align*}
\hfill \break
where we used the substitution $\theta_x=\arctan\left(\frac{k_x}{k}\right)$ and the Taylor series for $\arctan$ (which is absolutely convergent on $\mathbb{R}$).

\newpage
\noindent The appearing integral can be calculated analytically with the gamma-function $\Gamma$:

\begin{align*}
    &\int_\mathbb{R}k_x^{2n+1}\exp\left(-\frac{\left(k_x-k\frac{x}{R(z)}\right)^2}{2k_1^2(z)}\right)\,dk_x\\
    &=\int_\mathbb{R}\left(k_x+k\frac{x}{R(z)}\right)^{2n+1}\exp\left(-\frac{k_x^2}{2k_1^2(z)}\right)\,dk_x\\
    &=\int_\mathbb{R}\sum_{m=0}^{2n+1}\binom{2n+1}{m}k_x^m\left(k\frac{x}{R(z)}\right)^{2n+1-m}\exp\left(-\frac{k_x^2}{2k_1^2(z)}\right)\,dk_x\\
    &=\sum_{m=0}^{2n+1}\binom{2n+1}{m}\left(k\frac{x}{R(z)}\right)^{2n+1-m}k_1^m(z)\int_\mathbb{R}\left(\frac{k_x}{k_1(z)}\right)^m\exp\left(-\frac{\left(\frac{k_x}{k_1(z)}\right)^2}{2}\right)\,dk_x\\
    &=\sum_{m=0}^{2n+1}\binom{2n+1}{m}\left(k\frac{x}{R(z)}\right)^{2n+1-m}k_1^{m+1}(z)\int_\mathbb{R}u^m\exp\left(-\frac{u^2}{2}\right)\,dk_x\\
    &=\sum_{m=0}^{2n+1}\binom{2n+1}{m}\left(k\frac{x}{R(z)}\right)^{2n+1-m}k_1^{m+1}(z)2^{\frac{m-1}{2}}(1+(-1)^m)\Gamma\left(\frac{m+1}{2}\right)\\
    &=\sum_{l=0}^{2n+1}\binom{2n+1}{2l}\left(k\frac{x}{R(z)}\right)^{2n+1-2l}k_1^{2l+1}(z)2^{\frac{2l+1}{2}}\Gamma\left(\frac{2l+1}{2}\right)
\end{align*}
\hfill \break
Here we used the binomial expansion for $(a+b)^N$. From this follows:

\begin{align}
    &\mathbb{E}_{\theta_x}(Y)=\frac{1}{\sqrt{\pi}}\sum_{n=0}^\infty\frac{(-1)^n}{2n+1}\sum_{l=0}^n\binom{2n+1}{2l}\left( \frac{k}{k_1(z)}\right)^{2l}\left(\frac{x}{R(z)}\right)^{2(n-l)+1}2^l\Gamma\left(\frac{2l+1}{2}\right)\label{eq:EY}\\
    &\Rightarrow \mathbb{E}_x^2(\mathbb{E}_{\theta_x}(Y))=0\label{eq:E2EY}
\end{align}
\hfill \break
because $x^{2(n-l)+1}$ is an odd function and $f_x$ is an even function. From Eq.~\ref{eq:EY} one can write:

\begin{align*}
    \mathbb{E}_{\theta_x}(Y)&=\sum_{N=0}^\infty a_Nx^{2N+1}\\
    \Rightarrow \mathbb{E}_{\theta_x}^2(Y)&=\sum_{N=1}^\infty b_Nx^{2N}
\end{align*}

\newpage
Similarly to the calculation above one can interchange the integral with the power series and needs to calculate:

\begin{align*}
    \mathbb{E}_x\left(x^{2N}\right)&=\int_\mathbb{R}x^{2N}\exp\left(-2\frac{x^2}{w^2(z)}\right)\sqrt{\frac{2}{\pi w^2(z)}}\,dx\\
    &=\sqrt{\frac{2}{\pi w^2(z)}}\int_\mathbb{R}\left(\frac{2x}{w(z)}\right)^{2N}\left(\frac{w(z)}{2}\right)^{2N}\exp\left(-\frac{\left(\frac{2x}{w(z)}\right)^2}{2}\right)\,dx\\
    &=\sqrt{\frac{2}{\pi w^2(z)}}\left(\frac{w(z)}{2}\right)^{2N}\int_\mathbb{R}u^{2N}\exp\left(-\frac{u^2}{2}\right)\frac{w(z)}{2}\,du\\
    &=\sqrt{\frac{2}{\pi w^2(z)}}\left(\frac{w(z)}{2}\right)^{2N+1}2^{\frac{2N-1}{2}}(1+(-1)^{2N})\Gamma\left(\frac{2N+1}{2}\right)\\
    &=\frac{1}{\sqrt{\pi}}\left(\frac{w^2(z)}{2}\right)^N\Gamma\left(\frac{2N+1}{2}\right)
\end{align*}
\begin{equation}
    \Rightarrow \mathbb{E}_x(\mathbb{E}_{\theta_x}^2(Y))=\frac{1}{\sqrt{\pi}}\sum_{N=1}^\infty b_N\left(\frac{w^2(z)}{2}\right)^N\Gamma\left(\frac{2N+1}{2}\right) \label{eq:EE2Y}
\end{equation}
\hfill \break
Now only $\mathbb{E}(Y^2)$ is missing for Eqs.~\ref{eq:nominator} and \ref{eq:denominator}. For this define $c_n$ by:

\begin{equation}\label{eq:c2k}
    \arctan^2(x)=\sum_{n=2}^\infty c_nx^n\Rightarrow c_{2k}=\sum_{m=0}^{k-1}\frac{(-1)^{k-1}}{(2m+1)(2(k-m)-1)}    
\end{equation}
\hfill \break
Then one can calculate $\mathbb{E}_{\theta_x}(Y^2)$ similarly to $\mathbb{E}_{\theta_x}(Y)$:

\begin{align*}
    \mathbb{E}_{\theta_x}\left(Y^2\right)&=\int_\mathbb{R}\frac{\arctan^2\left(\frac{k_x}{k}\right)}{\sqrt{2\pi}k_1(z)}\exp\left(-\frac{\left(k_x-k\frac{x}{R(z)}\right)^2}{2k_1^2(z)}\right)\,dk_x\\
    &=\sum_{n=2}^\infty\frac{c_n}{\sqrt{2\pi}k_1(z)k^n}\int_\mathbb{R}k_x^n\exp\left(-\frac{\left(k_x-k\frac{x}{R(z)}\right)^2}{2k_1^2(z)}\right)\,dk_x\\
    &=\sum_{n=2}^\infty\frac{c_n}{\sqrt{2\pi}k_1(z)k^n}\sum_{m=0}^n\binom{n}{m}\left(k\frac{x}{R(z)}\right)^{n-m}k_1^{m+1}(z)2^{\frac{m-1}{2}}(1+(-1)^m)\Gamma\left(\frac{m+1}{2}\right)\\
    &=\sum_{n=2}^\infty\frac{c_n}{\sqrt{2\pi}}\sum_{m=0}^n\binom{n}{m}\left(\frac{k_1(z)}{k}\right)^m\left(\frac{x}{R(z)}\right)^{n-m}2^{\frac{m-1}{2}}(1+(-1)^m)\Gamma\left(\frac{m+1}{2}\right)
\end{align*}
\hfill \break
The expected value:

\begin{equation*}
    \mathbb{E}_x\left(\left(\frac{x}{R(z)}\right)^{n-m}\right)=\frac{1}{\sqrt{\pi}}\left(\frac{w(z)}{2R(z)}\right)^{n-m}2^{\frac{n-m-2}{2}}(1-(-1)^{n-m})\Gamma\left(\frac{n-m+1}{2}\right)    
\end{equation*}
\hfill \break
finishes the calculation:

\begin{align*}
    &\mathbb{E}\left(Y^2\right)\\
    &=\mathbb{E}_x(\mathbb{E}_{\theta_x}(Y^2))\\
    &=\frac{1}{\pi}\sum_{n=2}^\infty c_n\sum_ {m=0}^n\binom{n}{m}\left(\frac{k_1(z)}{k}\right)^m2^\frac{m}{2}\frac{1+(-1)^m}{2}\Gamma\left(\frac{m+1}{2}\right)\left(\frac{w(z)}{2R(z)}\right)^{n-m}2^{\frac{n-m}{2}}\frac{1+(-1)^{n-m}}{2}\Gamma\left(\frac{n-m+1}{2}\right)\\
    &=\frac{1}{\pi}\sum_{n=2}^\infty c_n2^{-\frac{n}{2}}\sum_{l=0}^{\lfloor\frac{n}{2}\rfloor}\binom{n}{2l}\left(\frac{k_1(z)}{k}\right)^{2l}\Gamma\left(\frac{2l+1}{2}\right)\left(\frac{w(z)}{R(z)}\right)^{n-2l}\frac{1+(-1)^{n-2l}}{2}\Gamma\left(\frac{n-2l+1}{2}\right)2^{2l}\\
    &=\frac{1}{\pi}\sum_{k=1}^\infty c_{2k}\left(\frac{w^2(z)}{2R^2(z)}\right)^k\sum_{l=0}^k\binom{2k}{2l}\left(\frac{k_1(z)}{k}\right)^{2l}\left(\frac{w(z)}{R(z)}\right)^{-2l}\Gamma\left(\frac{2l+1}{2}\right)\Gamma\left(\frac{2(k-l)+1}{2}\right)2^{2l}\\
    &=\frac{1}{\pi}\sum_{k=1}^\infty \left(\sum_{m=0}^{k-1}\frac{(-1)^{k-1}}{(2m+1)(2(k-m)-1)}\right)\left(\frac{w^2(z)}{2R^2(z)}\right)^k\sum_{l=0}^k\binom{2k}{2l}\left(\frac{2k_1(z)R(z)}{kw(z)}\right)^{2l}\Gamma\left(\frac{2l+1}{2}\right)\Gamma\left(\frac{2(k-l)+1}{2}\right)\\
    &=\frac{1}{\pi}\sum_{k=1}^\infty\left(-\frac{w^2(z)}{2R^2(z)}\right)^k \left(\sum_{m=0}^{k-1}\frac{-1}{(2m+1)(2(k-m)-1)}\right)\sum_{l=0}^k\binom{2k}{2l}\left(\frac{2k_1(z)R(z)}{kw(z)}\right)^{2l}\Gamma\left(\frac{2l+1}{2}\right)\Gamma\left(\frac{2(k-l)+1}{2}\right)
\end{align*}
\hfill \break
where we used Eq.~\ref{eq:c2k} to calculate $c_{2k}$. With this formula and Eqs.~\ref{eq:nominator}, \ref{eq:denominator}, \ref{eq:E2EY}, \ref{eq:EE2Y} one can calculate expected value in Eq.~\ref{eq:R2}. The result in paraxial approximation is given by Eq.~\ref{eq:r2}.

\end{document}